\documentclass[a4,fleqn,usenatbib]{mnras}
\usepackage{etoolbox}
\makeatletter
\patchcmd\@combinedblfloats{\box\@outputbox}{\unvbox\@outputbox}{}{%
   \errmessage{\noexpand\@combinedblfloats could not be patched}%
}%
 \makeatother

\usepackage[T1]{fontenc}
\usepackage{ae,aecompl}

\usepackage{graphicx}	
\usepackage{amsmath}	
\usepackage{amssymb}	
\usepackage{stmaryrd}
\usepackage{bm}
\usepackage{times}
\usepackage{amsmath}
\usepackage{amssymb}
\usepackage{textcomp}
\usepackage{verbatim}

\newcommand {\hi} {\ifmmode \ion{H}{I} \else $\ion{H}{I}$\fi}
\newcommand {\hii} {\ifmmode \ion{H}{II} \else $\ion{H}{II}$\fi}
\newcommand {\hei} {\ifmmode \ion{He}{I} \else $\ion{He}{I}$\fi}
\newcommand {\heii} {\ifmmode \ion{He}{II} \else $\ion{He}{II}$\fi}
\newcommand {\heiii} {\ifmmode \ion{He}{III} \else $\ion{He}{III}$\fi}
\newcommand {\HM} {\ifmmode \ion{H}{$_2$} \else $\ion{H}{$_2$}$\fi}
\newcommand {\xh} {\ifmmode X_\text{H} \else $X_\text{H}$\fi}
\newcommand {\nh} {\ifmmode n_\text{H} \else $n_\text{H}$\fi}

\newcommand {\areport} {{\sc arepo-rt}}
\newcommand {\arepo} {{\sc arepo}}

\title[Simulating the ISM of galaxies]{Simulating the interstellar medium of galaxies with radiative transfer, non-equilibrium thermochemistry, and dust}
\author[R. Kannan et al.]{Rahul Kannan,$^{1}$\thanks{E-mail: rahul.kannan@cfa.harvard.edu}\thanks{Einstein Fellow} Federico Marinacci,$^2$ Mark Vogelsberger,$^3$ Laura V. Sales,$^4$ Paul Torrey,$^5$ 
\newauthor Volker Springel$^6$ and Lars Hernquist$^1$
\\ 
\\
$^{1}$Center for Astrophysics | Harvard $\&$ Smithsonian, 60 Garden Street, Cambridge 02138, MA, USA\\
$^2$Department of Physics $\&$ Astronomy, University of Bologna, via Gobetti 93/2, 40129 Bologna, Italy\\
$^3$Kavli Institute for Astrophysics and Space Research, Massachusetts Institute of Technology, Cambridge, MA 02139, USA\\
$^4$Department of Physics $\&$ Astronomy, University of California, Riverside, 900 University Avenue, Riverside, CA 92521, USA\\
$^5$Department of Astronomy, University of Florida, 211 Bryant Space Sciences Center, Gainesville, FL 32611 USA\\
$^6$Max-Planck-Institut f\"ur Astrophysik, Karl-Schwarzschild-Str. 1, D-85748, Garching, Germany\\
}

\date{Accepted XXX. Received YYY; in original form ZZZ}

\pubyear{2019}

\begin{document}
\label{firstpage}
\pagerange{\pageref{firstpage}--\pageref{lastpage}}
\maketitle

\begin{abstract}
We present a novel framework to self-consistently model the effects of radiation fields, dust physics and molecular chemistry (\HM) in the interstellar medium (ISM) of galaxies.  The model combines a state-of-the-art radiation hydrodynamics module with a non-equilibrium thermochemistry module that accounts for \HM~coupled to a realistic dust formation and destruction model, all integrated into the new stellar feedback framework  {\it SMUGGLE}. We test this model on high-resolution isolated Milky-Way (MW) simulations.  We show that photoheating from young stars makes stellar feedback more efficient, but this effect is quite modest in low gas surface density galaxies like the MW. The multi-phase structure of the ISM, however, is highly dependent on the strength of the interstellar radiation field. We are also able to predict the distribution of \HM,  that allow us to match the molecular Kennicutt-Schmidt (KS) relation, without calibrating for it. We show that the dust distribution is a complex function of density, temperature and ionization state of the gas which cannot be reproduced by simple scaling relations often used in the literature.  Our model is only able to match the observed dust temperature distribution if radiation from the old stellar population is considered, implying that these stars have a  non-negligible contribution to dust heating in the ISM.  Our state-of-the-art model is well-suited for performing next generation cosmological galaxy formation simulations, which will be able to predict a wide range of resolved ($\sim 10$~pc) properties of galaxies.
\end{abstract}


\section{Introduction}
\label{sec:intro}
The formation and evolution of galaxies is governed by a complex interplay between a variety of physical processes such as gravity, gas cooling, star formation, stellar and black hole feedback, radiation, magnetic fields, cosmic rays etc. \citep[see for e.g., ][]{Benson2010, Naab2017, VN2019}. This implies that an ab initio model for galaxy formation would require self-consistent modelling of all these mechanisms. Unfortunately, the scales at which these processes operate are so disparate that it becomes impossible to model all of them properly at the same time. For example, gas inflows into the galaxy are determined by the large scale ($\sim$~Mpc) matter distribution of the Universe \citep{Freeke2011}, gas cooling rates depend on local gas densities, temperatures and metallicities \citep{Wiersma2009}, which in turn depend on the star formation and metal entrainment processes in galactic winds. These winds are launched close to star-forming regions, on scales of less than a parsec \citep{Thornton1998}. The effect of radiation fields is mainly felt around newly formed stars \citep[$\sim 1$~pc; ][]{Kannan2018} while black hole feeding and feedback require the simulations to resolve the accretion disc around the black hole \citep[$\sim 10^{-3}$~pc; ][]{Morgan2010}. 

The large dynamical range necessitates the use of analytic prescriptions that translate effects occurring below the resolution level of the simulation onto the grid scale. The accuracy of these sub-resolution models has been constantly improving. Current state-of-the-art simulations are able to capture and match a wide array of galaxy properties such as the galaxy luminosity function, colour bimodality, galaxy sizes, metallicities, etc. \citep{Stinson2013, Kannan2014, Vogelsberger2014, Vogelsberger2014a, Schaye2015, Wang2015, Dave2016, Dubois2016, Grand2017, Springel2018, Pillepich2018, Nelson2018, Dave2019}.   While they are quite good at reproducing the global properties of galaxies, they do not predict the small-scale structure ($\lesssim 1$~kpc) of star-forming regions. Moreover, most of these models are tuned to reproduce the observables in simulations with relatively low resolution and, therefore, it is unclear if they are adequate to simulate galaxies in a resolved manner. 

In recent years, increases in computational power have made it possible to simulate galaxy formation in a more resolved manner. These models have prescriptions for low temperature gas cooling, supernova (SN) energy and momentum input based on high resolution simulations of SN explosions \citep[e.g., ][]{Thornton1998, Kim2015, Martizzi2015} and a stochastic model for photoheating and radiation pressure feedback \citep{Fire1, Fire2, Smith2018, smuggle}.  The resulting simulations partially resolve giant molecular clouds, which are the sites for star formation, produce a multiphase interstellar medium (ISM) in a self-consistent manner, and reproduce resolved properties of galaxies such as local group dwarfs \citep{Wetzel2016, Smith2019}, globular cluster formation \citep{Ma2019}, and disc morphology and kinematics \citep{Ma2017}.  

While these models have introduced more realism into galaxy formation simulations, they still miss important processes like radiation fields, dust, molecular chemistry, and cosmic rays.  Radiation field can affect the properties of gas in galaxies on both large and small scales.  The intense radiation from newly formed young stars photoheats the surrounding high-density gas to a temperature of about $10^4$~K \citep{Stromgren1939}, driving small scale winds that reduce the density around stars before they explode, increasing the efficiency of SN feedback \citep{Rosdahl2015, Geen2015, Kannan2018}. Radiation escaping these star-forming clouds will ionize the gas and metals in the ISM and circumgalctic medium (CGM), lowering the cooling rates, in turn reducing the star formation rate in galaxies \citep{Cantalupo2010, Kannan2014a, Kannan2016}. Far-UV radiation can knock electrons off the dust grains in the ISM,  which thermalize and heat the gas \citep{Draine1978}. Radiation pressure, both by UV and trapped IR radiation can impart momentum into the ISM, which helps drive large scale galactic winds \citep{Murray2011}. While some sub-resolution models for the effect of radiation exists \citep{Fire1, Hu2017, Fire2, smuggle}, it is unclear if they accurately capture this process in a wide variety of environments \citep{Rosdahl2015, Kannan2018, Fujimoto2019}.

Dust is another important component of the ISM. Interstellar dust grains are microscopic solid particles made up from metals in the ISM. Although they account for only about $1\%$ of the total ISM mass they reprocess a large fraction of starlight, reradiating  it in the infra-red \citep[see, e.g., ][]{Clements1996}, are responsible for heating the low-density gas in the ISM via the photoelectric effect, and act as catalysts for numerous chemical reactions, including the formation of molecular hydrogen \citep[\ion{H}{$_2$, }][]{Gould1963}. It is important to understand dust abundances and obscuration in order to properly quantify star formation and stellar properties of galaxies \citep{Cochrane2019, Vogelsberger2019}.  Dust also determines the escape fraction of Lyman continuum photons from star forming clouds \citep{Kim2019} and is essential in driving outflows in highly optically thick galaxies \citep{Davis2014}. Finally, molecular hydrogen is an important cooling channel in the ISM, and the star formation rate of galaxies correlates extremely well with the \HM~content even on sub-kpc scales \citep{Leroy2008}. 

While these processes have been studied individually in galaxies \citep[e.g., ][]{Glover2007, Gnedin2011, Chrsitensen2012, Asano2013, Rosdahl2015, McKinnon2016, Richings2016,  McKinnon2017, HopkinsRT, McKinnon2018, Agertz2019, Nickerson2019}, their combined impact has still not been modeled properly.  This paper combines a state-of-the-art radiation hydrodynamics module with a non-equilibrium thermochemistry module that accounts for \HM~coupled to realistic dust formation and destruction, all integrated into a novel stellar feedback framework, the Stars and Multiphase Gas in Galaxies ({\it SMUGGLE}) model \citep{smuggle}. We test this model on isolated MW simulation at different resolutions and radiation field properties. We show that we are able to reproduce a wide range of observables of local group galaxies.  In Section~\ref{sec:methods} we outline the methods used in the paper. Section~\ref{sec:results} gives the most important results of this work and finally, Section~\ref{sec:conc}, summarizes our findings and outlines the main conclusions. 

\section{Methods}
\label{sec:methods}
The simulations presented in this work are performed with \areport  \  \citep{Kannan2019}, a radiation hydrodynamic (RHD) extension to the moving mesh hydrodynamic code {\sc arepo} \citep{Springel2010, Pakmor2016}. The mesh is regularized using the scheme described in \citet{Vogelsberger2012}.  The gravitational forces are calculated using a hierarchical octtree algorithm \citep{Barnes1986}. Radiation fields are simulated by casting the radiative transfer equation into a set of hyperbolic conservation laws for photon number density (${N}_\gamma$) and photon flux ({\bf F}$_\gamma$) by taking its zeroth and first moments respectively \citep{Kannan2019}. These equations are closed using the M1 closure relation \citep{Levermore1984}. The algorithm is fully conservative and compatible with the individual timestepping scheme of \arepo. In order to prevent extremely small timesteps, we use the reduced speed of light approximation \citep{Gnedin2001} with ${\tilde c} = 10^3 \ \mathrm{km \ s}^{-1}$.  Furthermore, for each hydro timestep the RT step is sub-cycled $8$ times according to the algorithm described in Appendix A of \citet{Kannan2019}. In the following sub-sections we will give a brief description of the gas cooling and thermochemistry, dust physics, star formation, radiative feedback, stellar winds and supernova (SN) feedback prescriptions used in this work.

\subsection{The thermochemical network and gas cooling }
\label{sec:cooling}
The cooling function is split into four separate terms: primordial cooling from Hydrogen and Helium ($\Lambda_p$), metal cooling ($\Lambda_M$), photoelectric heating ($\Lambda_\text{PE}$) and cooling due to dust-gas-radiation field ($\Lambda_D$) interactions.  The total cooling ($\Lambda_\mathrm{tot}$) is then given by:
\begin{equation}
\begin{split}
\Lambda_\text{tot} &= \Lambda_p(n_j, N_\gamma^i, T) + \frac{Z}{Z_\odot} \Lambda_M(T, \rho, z)\\ &+ \Lambda_\text{PE}(D, T, N_\gamma^\text{FUV}) + \Lambda_D(\rho, T, D, N_\gamma^\text{IR}) \, ,
\end{split} 
\end{equation}
where $T$ is the temperature of the gas, $Z$ is the metallicity, $Z_\odot$ is solar metallicity, $z$ is the redshift, $D$ is the dust to gas ratio, $\rho$ is the density of the gas cell, $n_j$ is the number density of the all the ionic species tracked in our primordial thermochemical network ($j \in [\HM, \hi, \hii, \hei, \heii, \heiii]$), and $N_\gamma^\text{FUV}$ and $N_\gamma^\text{IR}$ are the photon number density in the Infra-red (IR) and Far Ultra-Violet (FUV) bands, respectively (see Table~\ref{tab:rad} for more details). The primordial thermochemistry network couples the RHD module to the gas via photoheating and radiation pressure. 

In this work, we add a simple model for molecular Hydrogen (\HM) chemistry and cooling in addition to the atomic Hydrogen and Helium thermochemistry network described by Eqs.~$49-51$ of \citet{Kannan2019}. An additional equation describing the number density evolution of \HM~is incorporated into the chemical network:
\begin{equation}
\begin{split}
\frac{\text{d} n_\HM}{\text{d}t} &= \alpha_\HM^D \, \left(\frac{D}{D_\text{MW}}\right) \, n_{\text H} \, n_\hi + \alpha_\HM^\text{GP} \, n_\hi \, n_e \\ 
&+  \alpha_\HM^\text{3B} \, n_\hi^2 \, (n_\hi + n_\HM/8)  - \sigma_{\HM\hi} \,  n_\HM \, n_\hi \\ 
&- \sigma_{\HM\HM} \,  n_\HM^2  -   \, n_\HM \left( {\text S}_\HM  \Gamma_{\HM}^\text{LW} + \Gamma_{\HM}^+ \right) \, .
\label{eq:H2}
\end{split}
\end{equation}  
This also changes Eq.~52 of \citet{Kannan2019} to
\begin{equation}
\nh = 2n_\HM + n_\hi + n_\hii \, ,
\end{equation}
and all other relevant equations remain the same. 

In Eq.~\ref{eq:H2}, $\alpha_\HM^D$ is the formation rate of molecular hydrogen on dust grains, $D$ is the dust-to-gas ratio (modelled self-consistently using the dust model described in Section~\ref{sec:dust}), $D_\text{MW}$ is the dust-to-gas ratio in the Milky-Way (MW; $0.01$), $\alpha_\HM^\text{GP}$ is the formation rate of molecular hydrogen in the gas phase, $\alpha_\HM^\text{3B}$ is the formation rate through three body interactions, $\sigma_{\HM\hi}$ and $\sigma_{\HM \HM}$ are the collisional destruction rates due to collisions between \hi~ and \HM, and \HM~and \HM~respectively, $\Gamma_{\HM}^\text{LW}$ is the photodissociation rate due to Lyman-Werner band ($11.2-13.6 \ \text{eV}$) photons, and $ \Gamma_{\HM}^+$ is the photoionization rate due to photons with energies greater than $15.2 \ \text{eV}$. The value of all these coefficients are taken from \citet{Nickerson2018}. ${\text S}_\HM$ is the shielding factor due to molecular hydrogen \citep[see Eq. A5 of][]{Gnedin2011}. We follow \citet{Draine1996} and define this factor as
\begin{equation}
{\text S}_\HM = \frac{1-\omega_\HM}{(1+\chi)^2} + \frac{\omega_\HM}{\sqrt{1+\chi}} e^{-0.00085\sqrt{1+\chi}} \ ,
\end{equation}
where $\omega_\HM=0.2$ and $\chi = N_\HM/(5 \times 10^{14} \ \text{cm}^{-2})$.  The column density N$_\HM$ is obtained by using the Sobolev approximation $\text{N}_\HM = n_\HM L_\mathrm{sob}$, where $L_\mathrm{sob} = \rho/2|{\bf \nabla} \rho|$. The self-shielding factor is necessary because our RT implementation does not capture the self-shielding due to line overlap \citep[see ][ for more details]{Draine1996}. 

The cooling/heating rates for atomic Hydrogen and Helium are taken from \citet{Katz1996} and \citet{Kannan2019}.  The cooling due to molecular hydrogen is given by
\begin{equation}
\Lambda({\HM}) = \Lambda({n \to 0})_{\HM \hi} \, n_\HM \, n_\hi  + \Lambda({n \to 0})_{\HM \HM} \, n_\HM^2 \, ,
\end{equation}
where $\Lambda({n \to 0})_{\HM \hi}$ and $\Lambda({n \to 0})_{\HM \HM}$ are the low-density limits
of the \HM~collisional cooling coefficients as described in \citet{Hollenbach1979}. We ignore UV pumping heating and \HM~formation heating as they are sub-dominant in the regimes we consider \citep{Nickerson2018}. 
Cosmic ray ionization and heating are included in a very crude manner by assuming canonical MW ionization and heating rates of 
\begin{equation}
\begin{split}
&\Gamma_\HM^\text{cr} = 7.525 \times 10^{-16} \ \text{s}^{-1} \, , \\
&\Gamma_\hi^\text{cr} = 4.45 \times 10^{-16} \ \text{s}^{-1} \, , \\
&\Gamma_\hei^\text{cr} = 1.1  \Gamma_\hi^\text{cr} \ \text{and} \\
&\Lambda_\text{cr} = -1.6022 \times 10^{-11}  (\Gamma_j^\text{cr} \, n_j) \ \text{erg \ cm}^{-3} \ \text{s}^{-1} \, ,
\end{split}
\end{equation}
where $j \in [\HM, \hi, \hei]$. 

Metal-line cooling is implemented assuming ionization equilibrium for a given portion of dust-free and optically
thin gas in a UV background radiation field given by \citet{FG09}. The cooling rate
$\Lambda_\text{M}$ is computed from a look-up table containing the pre-calculated cooling values computed from CLOUDY  \citep[see][for more details]{Vogelsberger2013}. 

Photoelectric heating due to Far Ultra-Violet (FUV; $5.8 \ \text{eV} - 11.2 \ \text{eV}$) photons knocking off electrons from dust grains has been shown to be an important source of heating in the interstellar medium of galaxies \citep[see for eg.,][]{Bialy2019}. This heating rate is given by \citep{Wolfire2003}:
\begin{equation}
\Lambda_\text{PE} = - 1.3 \times 10^{-24} \, \left(\frac{\mathrm{D}}{\text{D}_\text{MW}}\right) \epsilon_\mathrm{ff} G \nh \,  \ \text{erg \ cm}^{-3} \ \text{s}^{-1} \,  ,
\label{eq:lambdape}
\end{equation}
where `$G$' is the flux if the FUV band normalized to $1.7$ times the Habing value \citep[$ G_0 = 1.6 \times 10^{-3} \ \text{erg \ s}^{-1} \ \text{cm}^{-2}$; ][]{Habing1968} and 
\begin{equation}
\begin{split}
\epsilon_\mathrm{ff}  &= \frac{4.87 \times 10^{-2}}{1+4 \times 10^{-3} (G_0 \sqrt{T}/(0.5 n_e))^{0.73}}\\ &+ \frac{3.65 \times 10^{-2} (T/10^4)^{0.7}}{1+2 \times 10^{-4} (G_0 \sqrt{T}/(0.5 n_e))} \, .
\end{split} 
\end{equation} 

The final cooling term we implement is that due to dust-gas energy ($\Lambda_\mathrm{gd}$) exchange via collisions. The rate of dust-gas energy exchange is given by \citep{Burke1983}
\begin{equation}
\Lambda_\mathrm{gd} = n_\mathrm{gr} \, n_\text{H} \, \sigma_\mathrm{gr} \, \overline{\alpha_T} \, 2k_B \,  \sqrt{\frac{8 k_B T}{\pi m_p}} \, (T - T_d) \, ,
\label{eq:lambdagd}
\end{equation}
where $n_\mathrm{gr}$ is the number density of dust grains, $\sigma_\mathrm{gr}$ is the collisional cross section of the dust grains, $T$ is the gas temperature, $T_d$ is the dust temperature and $\overline{\alpha_T}$ is the "average accommodation coefficient". If we assume that the dust grains have size `$a$' with density $\rho_\mathrm{gr}$ and the hydrogen mass fraction is \xh, then Eq.~\ref{eq:lambdagd} can be rewritten as
\begin{equation}
\Lambda_\mathrm{gd} = \left( \frac{3\rho_d}{4\pi\rho_\mathrm{gr} a^3} \right) \, \left(\frac{\rho_g \xh}{m_p}\right) \, (\pi a^2) \, \overline{\alpha_T} \, 2k_B \,  \sqrt{\frac{8 k_B T}{\pi m_p}} \, (T - T_d) \, ,
\end{equation}
by setting $\rho_d = D\rho_g$, $\rho_\mathrm{gr} = 2.4 \ \text{g \ cm}^{-3}$, $\xh = 0.76$ and $\overline{\alpha_T} = 0.5$ (see Figure~4 of \citealt{Burke1983}) and rearranging the terms we get
\begin{equation}
\begin{split}
\Lambda_{gd} &= \beta \left( \frac{D}{0.01}\right) \left( \frac{0.1 \mu m}{a}\right) \, \sqrt{\frac{T}{1 \ \text{K}}} \, \left(\frac{T - T_d}{1 \ \text{K}}\right) \, \left(\frac{n_H}{1 \ \text{cm}^{-3}}\right)^2 \, ,
\end{split}
\end{equation}
where
\begin{equation}
\begin{split}
\beta &= \left(\frac{3 \, m_p \, \overline{\alpha_T} \, 2 k_B}{4 \rho_\mathrm{gr} \,  \xh}\right) \sqrt{\frac{8 k_B}{\pi m_p}} \left(\frac{0.01}{0.1 \mu m}\right) \, \left( \frac{(1 \ \text{K})^{3/2}}{(1 \ \text{cm}^{3})^2} \right) \\
& = 1.356 \times 10^{-33}  \ \text{erg \ cm}^{-3} \ \text{s}^{-1} \, .
\end{split}
\label{eq:lambdad}
\end{equation}
The two unknowns in this equation are the dust-to-gas ratio ($D$) and the dust temperature ($T_d$) which are self consistently calculated from the empirical dust model described in the next section. We note that this cooling channel becomes relevant only in extreme high-density regions, which in our model are already star-forming.
\subsection{Dust Physics}
\label{sec:dust}
In this section we briefly describe the self-consistent dust formation and destruction model used in our simulations \citep[see][for more details]{McKinnon2017}. The model tracks the dust mass for five chemical species (C, O, Mg, Si and Fe) for each gas cell. The size of the dust grains is assumed to be constant throughout the simulation ($a = 0.1 \,  \mu \mathrm{m}$).  The dust is assumed to be dynamically coupled to the gas and is passively advected along with it \citep[see][for a live dust implementation]{McKinnon2018}. The model accounts for three distinct dust production channels namely, SNII, SN Ia and asymptotic giant branch (AGB) stars. The dust produced during the mass return from these stars follows the prescriptions outlined in \citet{Dwek1998}. 

The mass of dust in the ISM  increases due to the gas-phase metals depositing onto existing grains which is modeled according to the prescription of \citet{Dwek1998} and \citet{Hirashita1999}: 
\begin{equation}
 \frac{\text{d} M_\text{dust}}{\text{d}t} = \left( 1 - \frac{M_\text{dust}}{M_\text{metal}}\right) \left(\frac{M_\text{dust}}{\tau_g} \right) \, ,
\end{equation}
where $M_\text{dust}$ and $M_\text{metal}$ are the total mass of dust and metals in the cell, and $\tau_g$ is the characteristic dust growth timescale. This timescale depends on the density and temperature of the gas and is given by \citep{Yozin2014, Zhukovska2014}:
\begin{equation}
 \tau_g = \tau_g^\text{ref} \left( \frac{\rho^\text{ref}}{\rho} \right) \sqrt{\frac{T^\text{ref}}{T}} \, ,
\end{equation}
where $\tau_g^\text{ref}$ is a normalization which depends on atom-grain collision sticking efficiencies and grain cross-sections, and $\rho^\text{ref}$ and $T^\text{ref}$ are the reference density and temperature set to $1$~H~atom~$\text{cm}^{-3}$ and $20 \ \text{K}$ respectively. 

Dust is destroyed through shocks from SN remnants \citep[for eg.,][]{Seab1983} and thermal sputtering \citep[for eg.,][]{Draine1979}. 
\begin{equation}
  \frac{\text{d} M_\text{dust}}{\text{d}t} = -\frac{M_\text{dust}}{\tau_d} - \frac{3 \, M_\text{dust}}{\tau_\text{sp}} \, ,
\end{equation}
where $\tau_d$ is the dust destruction timescale due to SN shocks and is given by
\begin{equation}
 \tau_d = \frac{m_\text{gas}}{\beta \eta M_s(100)} \, ,
 \label{eq:SNdust}
\end{equation}
 where $m_\text{gas}$ is the mass of the cell,  $\eta$ is the local Type II SN rate, $\beta$ is the grain destruction efficiency in SN shocks, and $M_s(100)$ is  the amount of gas mass accelerated to at  least $100 \ \text{km \ s}^{-1}$ \citep{McKee1989}. Recent results suggest that $M_s(100)$ and $\beta$ will depend on the SN rate, environment and the properties and size of the dust grains \citep{dust19, Zhu19}. However, more work needs to be done in order to quantify the dependence of these parameters in a variety of environments.  Therefore, we use the simple relation (which has been shown to reproduce a wide variety of dust properties; \citealt{McKinnon2016, McKinnon2017})  in this work, with the goal to improve it once the physics is better understood. 
 
  The sputtering timescale is given by \citep{Tsai1995}
 \begin{equation}
  \tau_\text{sp} = (0.17 \ \text{Gyr}) \left(\frac{a}{0.1 \mu m}\right) \left( \frac{10^{-27} \ \text{g cm}^{-3}}{\rho}\right) \left[ \left(\frac{T_0}{T}\right)^\omega + 1 \right] \, ,
  \label{eq:dustsputter}
 \end{equation}
where $T_0 = 2 \times 10^6 \ \text{K}$ is the temperature above which the sputtering rate is constant and $\omega=2.5$ controls the fall off in the sputtering rate at low temperatures. 

In this work we additionally include a model to track the temperature of dust grains. In the previous section we outlined the equations governing the energy exchange between gas and dust due to collisions (Eq.~\ref{eq:lambdad}). The other important process of energy exchange is between the dust and the Infra-red (IR, $0.1 - 1 \ \text{eV}$) radiation field. This energy exchange rate is given by \citep{Kannan2019}:
\begin{equation}
 \Lambda_\text{dr} = \kappa_P \, \rho \, c \, ( aT_d^4 - E_\text{IR}) \, ,
\end{equation}
where $\kappa_P$ is the Planck mean opacity, $T_d$ is the dust temperature, $c$ is the speed of light, $a$ is the radiation constant  and $E_\text{IR}$ is the energy density of photons in the IR bin. $\kappa_P$ is calculated based on the local dust properties of the cell as outlined in Appendix C of \citet{Kannan2019}. Since dust grains reach thermal equilibrium on a rapid time-scale \citep{Woitke2006}, we calculate the dust temperature $T_d$ by solving the instantaneous equilibrium condition $\Lambda_\text{gd} + \Lambda_\text{dr} = 0 $ using Newton's method for root-finding. 
\subsection{Star formation}
Cold gas in our simulations is converted to star particles using the usual probabilistic approach outlined in \citet{Springel2003}. Stars are assumed to form only above a density threshold of  $n_\text{th} = 10^3 \, \mathrm{cm}^{-3}$.
The star formation efficiency is a free parameter which is set to $\epsilon=0.01$, in line with recent observational estimates \citep{Krumholz2007}. Additionally, following the {\it SMUGGLE} implementation, we impose the condition that the star forming gas cloud needs to be self-gravitating in order to form stars \citep[Equation~9; ][]{smuggle}. 
Although we do follow the formation of molecular hydrogen in our simulations,  we choose not to tie the star formation rate to the abundance of \HM, with the goal of recovering the molecular Kennicutt-Schmidt \citep{Bigiel2008, Leroy2008} relation naturally in our simulations without the need to impose it. 

Finally, a Jeans pressure floor is enforced in regions which are below the resolution limit of the simulation. This is done in order to prevent artificial numerical fragmentation in highly dense and underresolved regions. Specifically, the gas pressure of the cells ($P_\mathrm{cell}$) is set to
\begin{equation}
P_\mathrm{cell} = \mathrm{max}\left(P_\mathrm{hydro}, \frac{2 \, (\gamma -1) \, \rho_\mathrm{cell} \, G \, M_\mathrm{cell}}{\epsilon_\mathrm{gas}} \right) \, ,
\end{equation}
where $P_\mathrm{hydro}$ is the pressure obtained from the hydrodynamic solver, $\rho_\mathrm{cell}$ is the density of the cell, $G$ is the gravitational constant,  $\gamma = 5/3$ is the adiabatic index, $M_\mathrm{cell}$ is the mass of the cell, and $\epsilon_\mathrm{gas}$ is minimum softening length of gas particles. 

 \begin{table*}
\begin{center}
 \label{tab:example}
 \begin{tabular}{ccccccccccccccccc}
  \hline
 Bin & Range & $\sigma_{\HM}$ & $\sigma_{\hi}$ & $\sigma_{\hei}$ &$\sigma_{\heii}$ & $\mathfrak{h}_{\HM}$ & $\mathfrak{h}_{\hi}$ & $\mathfrak{h}_{\hei}$ & $\mathfrak{h}_{\heii}$ & $p_{\HM}$ & $p_{\hi}$ & $p_{\hei}$ &$p_{\heii}$ & $\mathcal{E}$ & $\kappa_d$ \\
 & [eV] & [Mb] & [Mb] & [Mb] & [Mb] & [eV] & [eV] & [eV] & [eV] & [eV] & [eV] & [eV] & [eV] & [eV] & [cm$^2$ g$^{-1}$] \\
  \hline
  IR & 0.1-1 & 0 & 0 & 0 & 0 & 0 & 0 & 0 & 0 &  0 & 0 & 0 & 0 & $0.65$ & - \\
  Opt & 1-11.2 & 0 & 0 & 0 & 0 & 0 & 0 & 0 & 0 &  0 & 0 & 0 & 0 & $6.40$ & 1000 \\
 LW & 11.2 -13.6 & 0.21 & 0 & 0 & 0 & 0 & 0 & 0 & 0 &  12.26 & 0 & 0 & 0 & $12.26$ & 1000 \\
 EUV1 & 13.6 -24.6 & 5.09 & 3.37 & 0 & 0 & 3.78 & 3.15 & 0 & 0 &  18.98 & 16.75 & 0 & 0 & $18.01$ & 1000 \\
 EUV2 & 24.6 - 54.4 & 2.42 & 0.79 & 5.10 & 0 & 14.75 & 13.10 & 3.81 & 0 &  28.35 & 28.30 & 28.41 & 0 & $29.89$ & 1000 \\
 EUV3 & 54.4 - $\infty$ & 0.32 & 0.11 & 0.77 & 1.42 & 41.18 & 42.78 & 31.80 & 2.03 &  56.38 & 56.38 & 56.40 & 56.43 & $56.85$ & 1000 \\
  \hline
 \end{tabular}
    \caption{Table outlining the frequency discretization of the radiation field as used in our simulations. It lists the frequency bin name (first column), the frequency range it covers (second column), the mean ionization cross section ($\sigma$, Eq.~\ref{eq:cross}) for the different species ({\HM,} third column; {\hi,} fourth column; {\hei,} fifth column; {\heii,} sixth column), the mean photoheating rate ($\mathfrak{h}$, Eq.~\ref{eq:photoheat}) for the different species ({\HM,} seventh column; {\hi,} eighth column; {\hei,} ninth column; {\heii,} tenth column), the mean radiation pressure ($p$, Eq.~\ref{eq:rp}) for the different species ({\HM,} eleventh column; {\hi,} twelfth column; {\hei,} thirteenth column; {\heii,} fourteenth column), the mean energy per photon ($\mathcal{E}$, fifteenth column), and the opacity to dust grains ($\kappa_d$, sixteenth column).}
     \label{tab:rad}

 \end{center}
\end{table*}

\subsection{Stellar feedback prescriptions}
We implement three feedback mechanisms related to young stars, namely, radiative feedback, stellar winds from young (O, B) stars and AGB stars and SN feedback. We retain the  SN and stellar wind feedback prescriptions of {\it SMUGGLE} and replace the sub-grid radiation feedback prescriptions for photoheating and radiation pressure with accurate radiation hydrodynamics. A brief description of the implementation of all three mechanisms is given below.

\subsubsection{Radiative feedback}
The main radiative feedback mechanisms, namely, photoheating, radiation pressure, and photoelectric heating are modeled self-consistently through the radiative transfer scheme. In our simulations, the newly formed star particles are a source of local radiation. The luminosity and spectral energy density is a complex function of age and metallicity taken from \citet[BC03;][]{Bruzual2003}. To increase the
probability of resolving the Str\"omgren radius, we inject all the photons in the nearest two cells
closest to the star particle. Additionally, the direction of the photon
flux (${\bf F}_\gamma$) is set to be radially outward from the star particle and the magnitude is $|{\bf F}_\gamma | = {\tilde c} E_\gamma$. This ensures that the full radiation pressure force is
accounted for even if the cell optical depth is larger than one \citep{Kannan2018}. 

We include photoinization and photoheating of \HM, \hi, \hei, and \heii\ species. The photoionization rate for each species `j' due to the radiation bin `i' is given by
\begin{equation}
\dot{n}_j = -\tilde{c} n_j \Sigma_i \overline{\sigma}_{ij} N_\gamma^i \, ,
\end{equation}
where $\overline{\sigma}_{ij}$ is the mean ionization cross section of photons in bin `i' for interacting with species `j' and is given by
\begin{equation}
{\bar \sigma_{ij}} = \frac{\displaystyle\int_{\nu_{i1}}^{\nu_{i2}} \frac{4\pi J_\nu}{h\nu} \sigma_{j_\nu} \, \text{d}\nu}{\displaystyle\int_{\nu_{i1}}^{\nu_{i2}}\displaystyle\frac{4\pi J_\nu}{h\nu} \, \text{d}\nu} \, .
\label{eq:cross}
\end{equation}
In this equation
\begin{equation}
J_\nu = \frac{1}{4\pi}\int_{4\pi} I_\nu \, \text{d}\Omega\, ,
\end{equation}
$\sigma_{j\nu}$ is the frequency dependent cross section of species `j', and $h$ is the Planck constant. 

Similarly, the total photoheating rate is given by
\begin{equation}
{\mathcal H} = \sum_j n_j \Gamma_j \, ,
\label{eq:UVheating}
\end{equation}
where 
\begin{equation}
 \Gamma_j =  {\tilde c} \sum_i  N_\gamma^i \, \bar{\sigma}_{ij} \, \mathfrak{h}_{ij} \, .
 \end{equation}
 The mean photoheating rate $\mathfrak{h}_{ij}$ is 
   \begin{equation}
\mathfrak{h}_{ij} = \frac{\displaystyle\int_{\nu_{i1}}^{\nu_{i2}} \frac{4\pi J_\nu}{h\nu} \, \sigma_{j_\nu} \, (h\nu-h\nu_{tj}) \, \text{d}\nu}{\displaystyle\int_{\nu_{i1}}^{\nu_{i2}} \frac{4\pi J_\nu}{h\nu} \, \sigma_{j_\nu} \, \text{d}\nu} \, ,
\label{eq:photoheat}
\end{equation}
where $\nu_{tj}$ is the threshold frequency for the ionization of species `j'. 

The radiation pressure term is added as a source term in the momentum conservation equation of hydrodynamics and is given by
\begin{equation}
\frac{\partial \rho v}{\partial t} =  \frac{1}{c} \sum_i {\bf F}_\gamma^i \, \left(\sum_j \, n_j \, \bar{\sigma}_{ij} \, p_{ij}  + \kappa_i \, \rho \, e_i\right)\, ,
\label{eq:uvmomentum}
\end{equation}
where
\begin{equation}
 p_{ij} = \frac{\displaystyle\int_{\nu_{i1}}^{\nu_{i2}} 4\pi J_\nu \, \sigma_{j_\nu} \, \text{d}\nu}{\displaystyle\int_{\nu_{i1}}^{\nu_{i2}} \frac{4\pi J_\nu}{h\nu} \, \sigma_{j_\nu} \, \text{d}\nu} \, ,
 \label{eq:rp}
\end{equation}
$\kappa_i$ is the opacity due to dust, and $e_i$ is the mean photon energy of bin `i'.  Moreover, the IR scattering scheme described in Section 3.2.2 of \citet{Kannan2019} automatically takes care of IR radiation pressure and the momentum boost at high optical depths, without the need for subgrid models. 

It is clear that the mean ionization cross section, photoheating rate, and the mean radiation pressure terms vary from cell to cell due to the differing shapes of the radiation spectrum from different age and metallicity sources.  The M1 scheme is not able to track this change in shape. We note that for the \citet{Bruzual2003} spectra the calculated radiation parameters are roughly constant and do not vary significantly with the metallicity  and age of the star \citep[see Figure~B2 of ][]{Rosdahl2013}. We therefore calculate them using the solar metallicity - $10$ Myr\ spectrum and use the same values for all the cells throughout the simulation. We discretise the radiation field in  six radiation bins, namely, the Infra-red (IR, $0.1-1$ eV) bin, the optical bin (Opt, $1-11.2$ eV), the Lyman-Warner band (LW, $11.2-13.6$ eV), hydrogen ionizing bin (EUV1, $13.6-24.6$ eV), \ion{He}{I} ionizing bin (EUV2, $24.6-54.4$ eV), and finally the \ion{He}{II} ionizing bin (EUV3, $54.4-\infty$ eV). The mean ionization cross section, photoheating rate, and mean radiation pressure for each of the bins are tabulated in Table~\ref{tab:rad}.

Finally, photoelectric heating is implemented using Eq.~(\ref{eq:lambdape}) described in Section~\ref{sec:cooling}. The flux of the radiation field in the FUV band is estimated in a self-consistent manner from out RT module. The frequency range of photons that instigate photoelectric heating is from $5.8-11.2 \ \text{eV}$.
\subsubsection{Stellar winds}
Stellar winds from young massive OB ($\gtrsim 8 \ \text{M}_\odot$) stars and the asymptotic giant branch (AGB) stars contribute significantly to stellar feedback. We include stellar winds according to the {\it SMUGGLE} model described in \citet{smuggle}. Briefly, the momentum input is computed in two parts. First, the mass loss from OB and AGB stars is calculated according to the analytic prescription given in \citet{Fire2}. The mass loss is quantified as a function of both the age of the star and its metallicity. 

The energy of the stellar wind is then computed as 
\begin{equation}
E_\mathrm{winds} = \delta t L_\mathrm{kin} = M_\mathrm{loss} \, \psi \, 10^{12} \ \text{erg g}^{-1} \, , 
\end{equation}
where $M_\mathrm{loss}$ is the mass loss rate and
\begin{equation}
\psi = \frac{5.94 \times 10^4}{1+\left(\frac{t}{2 \ \text{Myr}}\right)^{1.4} + \left(\frac{t}{10 \ \text{Myr}}\right)^5} + 4.83 \, .
\end{equation}
Therefore, the total momentum of the wind is 
\begin{equation}
p_\mathrm{winds} = \sqrt{2 \, M_\mathrm{loss} \, E_\mathrm{winds}} \, .
\end{equation}
The mass, momentum and energy from stellar winds are injected in a continuous manner in the rest frame of the star and then transformed back into
the reference frame of the simulations. 

\begin{table*}
\begin{center}
 \label{tab:example}
 \begin{tabular}{ccccccccccccc}
  \hline
 Galaxy & $M_{\text{halo}}$ & $v_{200}$ & c & $M_{\text{bulge}}$ & a & $M_{\text{disc}}$ & $r_d$ & $h$ & $M_\text{gas}$ &  $r_g$ & $f_\text{gas}$ & L$_\text{box}$\\
   & [M$_\odot$] & [km s$^{-1}$] & & [M$_\odot$] & [kpc] & [M$_\odot$] & [kpc] & [pc] & [M$_\odot$] &  [kpc] & [$R<R_\odot$] & [kpc]\\
  \hline
  MW & $1.53 \times 10^{12}$ & $169$ & 12 &  $1.5 \times 10^{10}$ & $1.0$ & $4.74 \times 10^{10}$ & $3.0$ & $300$ & $9 \times 10^9$ & 6.0 & 0.1 & 857 \\ 
  \hline
 \end{tabular}
  \caption{The parameters of the Milky-Way like disc used as the initial condition for the simulations.}
        \label{tab:param}
 \end{center}
\end{table*}

\begin{table*}
\begin{center}
 \label{tab:example}
 \begin{tabular}{ccccccccccccc}
  \hline
  Simulation & $m_{\star, \text{disc}}$  & $m_{\star, \text{bulge}}$ & $m_{\star, max}$ & $m_\text{gas}$  &  $\epsilon_\star$  & $\epsilon_\text{gas}$ & Model\\
    &  [M$_\odot$]  &  [M$_\odot$]  & [M$_\odot$] &  [M$_\odot$] &   [pc] &  [pc] & \\
  \hline
  MW-high & $1.9 \times 10^3$ & $2.3 \times 10^3$ & $2.8 \times 10^3$ & $1.4 \times 10^3$ & 7.1 & 3.6 & fiducial\\ 
    MW-low & $1.5  \times 10^4$ & $2 \times 10^4$ & $2.2 \times 10^4$& $1.1 \times 10^4$ & 21.4 & 10.7 & fiducial\\ 	
       MW-low-no-old & $1.5  \times 10^4$ & $2 \times 10^4$ & $2.2 \times 10^4$& $1.1 \times 10^4$ & 21.4 & 10.7 & no old stars\\
       MW-low-no-RT & $1.5  \times 10^4$ & $2 \times 10^4$ & $2.2 \times 10^4$& $1.1 \times 10^4$ & 21.4 & 10.7 & no radiation fields\\ 		
  \hline
 \end{tabular}
  \caption{Table of the simulations reported in this work and simulation parameters such as the model name (first column), mass of stars in the disc (second column), mass of stars in the bulge (third column),  maximum mass of newly formed stellar particles (fourth column), minimum mass of gas particles (fifth column), softening length of stars (sixth column) and stars (seventh column), and the model description (eighth column).}
      \label{tab:simulations}
 \end{center}
 \end{table*}
 
\begin{figure*}
\includegraphics[width=0.99\textwidth]{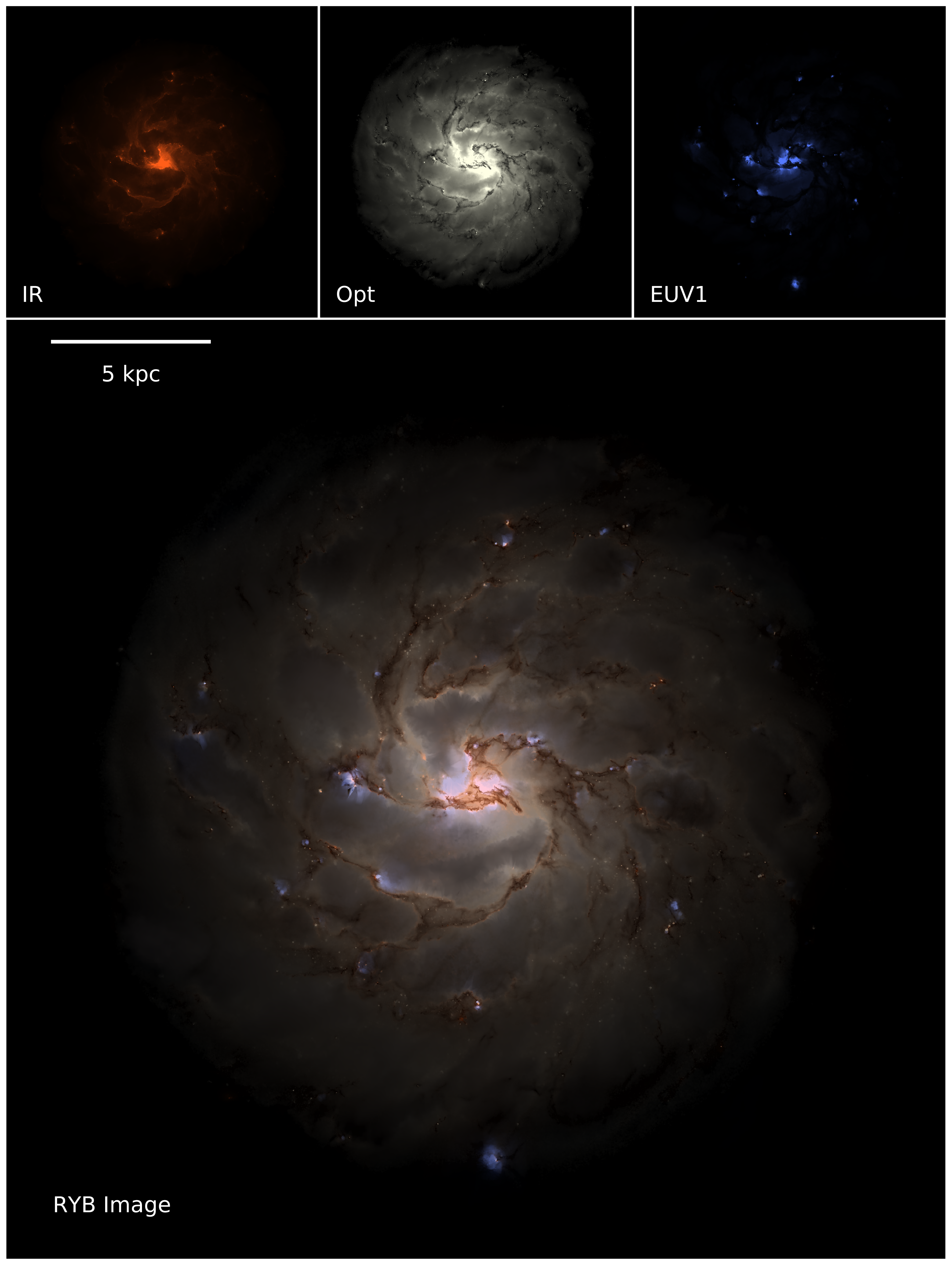}
\caption{IR (top left), Optical (top center), (top right) and false color RYB composite image (bottom panel) of the MW-high run after 1 Gyr of evolution.}
\label{fig:rgb}
\end{figure*}

\subsubsection{Supernova feedback}
Feedback from SN plays a crucial role in regulating star formation rate of low mass galaxies \citep[see for e.g., ][]{Kannan2014}.  As in \citet{smuggle}, we implement a boosted momentum injection method that takes into account the cooling loses that occur due to the inability to resolve the Sedov-Taylor phase.  The momentum input per SN event into a neighbouring cell `i' is 
\begin{equation}
\Delta p_i = \tilde{w_i} \, \text{min}\left[ p_\text{SN}\sqrt{1+\frac{m_i}{\Delta m_i}}, p_t\right] \, ,
\end{equation}
where $p_\text{SN} = \sqrt{2 M_\text{SN} E_\text{SN}}$ is the SN momentum carried by the SN event at the time of explosion, $\tilde{w}_i$ is a weight function dividing the energy and momentum injection, $m_i$ is the mass of the gas cell, $\Delta m_i$ is the mass released during the SN event, and $p_t$ is the terminal momentum, which is defined as the momentum of the SN blast when it transitions from the Sedov-Taylor phase to a momentum-conserving phase
\begin{equation}
p_t = 4.8 \times 10^5 \, E_\text{SN,tot}^{13/14} \left( \frac{\left< \nh \right>}{1 \ \text{cm}^{-3}}\right)^{-1/7} \left( \frac{\left< Z\right>}{Z_\odot}\right)^{-0.21} \ \text{M}_\odot \ \text{km s}^{-1} \, . 
\end{equation}
Here, $\left< \nh \right>$ and $\left< Z \right>$ are the average density and metallicity of the gas surrounding the star particle which  has undergone the SN event. Moreover, since SN explosions are discrete events, the model mimics their discrete nature by imposing a time-step constraint for each stellar particle based on its age, such that the expectation value for the number of SN events per timestep is of the order of unity. This form of SN feedback has been shown to regulate the star formation rate in low mass galaxies in a variety of environments \citep{Fire2, smuggle}.

\subsection{Initial Conditions}
We study the role of radiation fields, molecular chemistry, and dust physics in an isolated galaxy environment of a Milky-Way (MW) type galaxy (as in \citealt{smuggle}). We construct the initial conditions following the techniques described in \citet{Hernquist1993} and  \citet{Springel2005}. The initial galaxy is setup with a dark matter halo, a bulge, and stellar and gaseous discs. The DM halo and the bulge are modeled with a Hernquist profile \citep{Hernquist1990} with a scale length $a$. The gas and the stellar disc have an exponential profile in the radial direction with an effective radius $r_g$ and $r_d$ respectively. The vertical profile of the stellar disc follows a ${\rm sech}^2$ functional form with the scale height $h$. The vertical profile of the gaseous disc is computed assuming hydrostatic equilibrium, with the initial gas temperature set to $10^4$~K. The gas in the disc has a metallicity equal $Z = Z_\odot = 0.0127$ \citep{Asplund2009}. The lack of cosmological gas inflow into the disc can generate unrealistic gas metallicities. To avoid this we ensure that the gas and metal mass returned from the stars is always equal to the initial metallicity. The dark matter halo is modeled as a static background gravitational field, that is not impacted by the baryonic physics. All our simulations are evolved for $\sim 1 \ \text{Gyr}$ using the scheme described in Section~\ref{sec:methods}. The structural parameters of the galaxies under consideration are given in Table~\ref{tab:param}.

The galaxy setup is run at two different resolutions termed  'low' and 'high'. We run three low resolution runs, one containing the fiducial model, another the fiducial model without contribution to the radiation from old stars, defined as stars already present initially (we assign an age of $5$ Gyrs to all these stars), and finally the fiducial runs minus any local radiation fields. The single high resolution run employs the fiducial model. This set of simulations is used to understand the contribution of different radiation fields and also to test the convergence of our model. The simulations we perform and the respective parameters are tabulated in Table~\ref{tab:simulations}.

\begin{figure}
\includegraphics[width=\columnwidth]{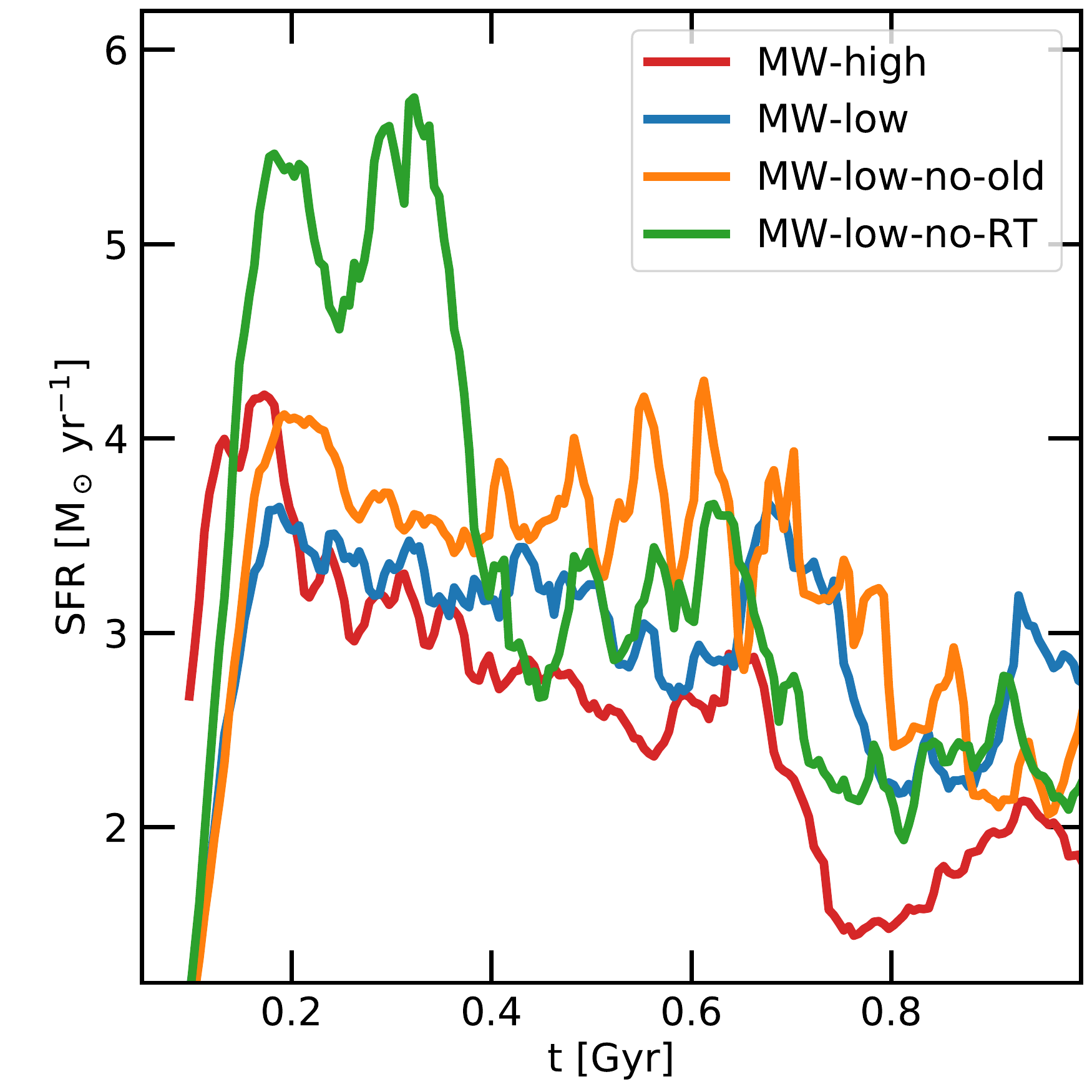}
\caption{The star formation rate history as a function of simulation time for the 'MW-high' (red curve), 'MW-low' (blue curve), 'MW-low-no-old' (orange curve), and 'MW-low-no-RT' (green curve) runs.}
\label{fig:SFR}
\end{figure}

\begin{figure*}
\includegraphics[width=\textwidth]{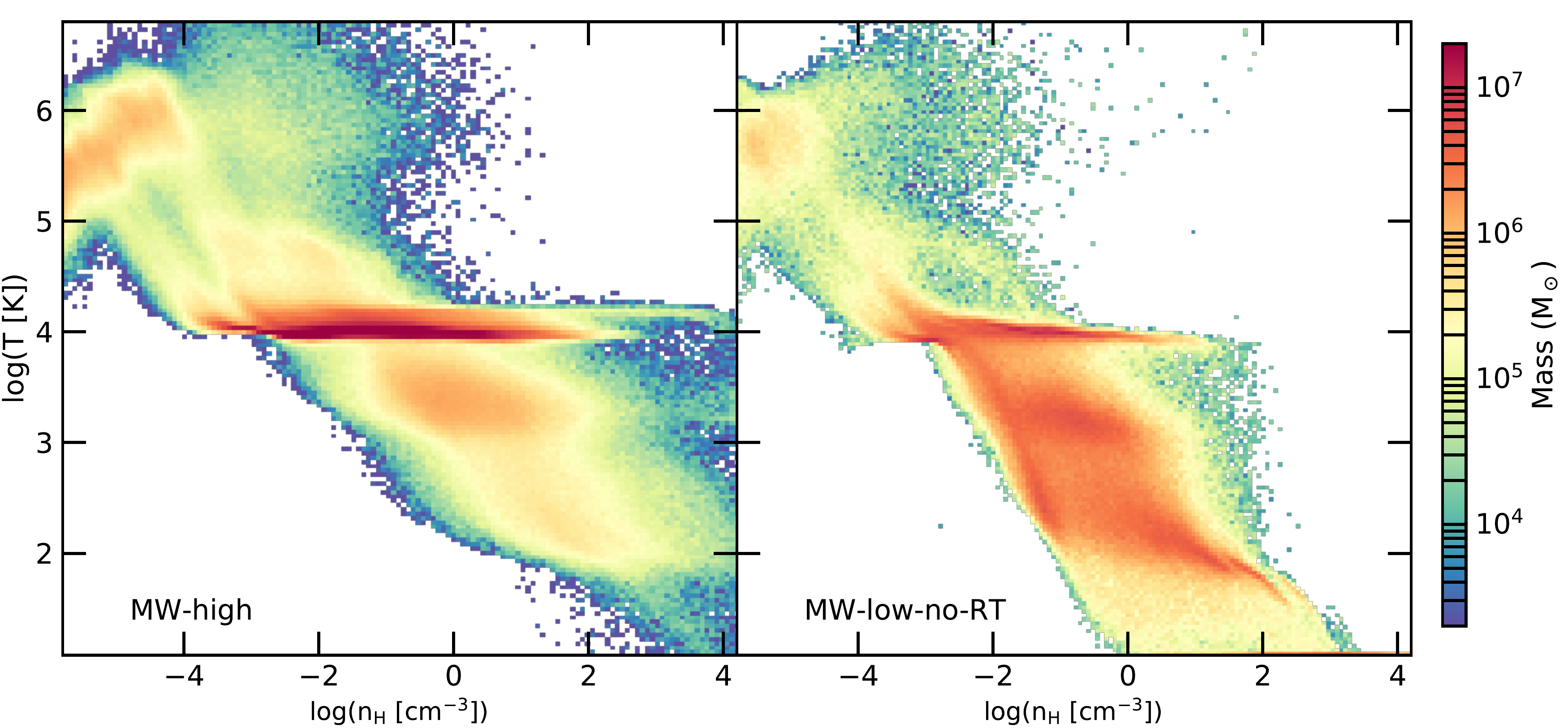}
\caption{The temperature - density phase space diagram of all gas, color coded by the gas mass in the MW-high (left panel) and MW-low-no-RT (right panel) simulations.}
\label{fig:phasespace}
\end{figure*}

\begin{figure*}
\includegraphics[width=\textwidth]{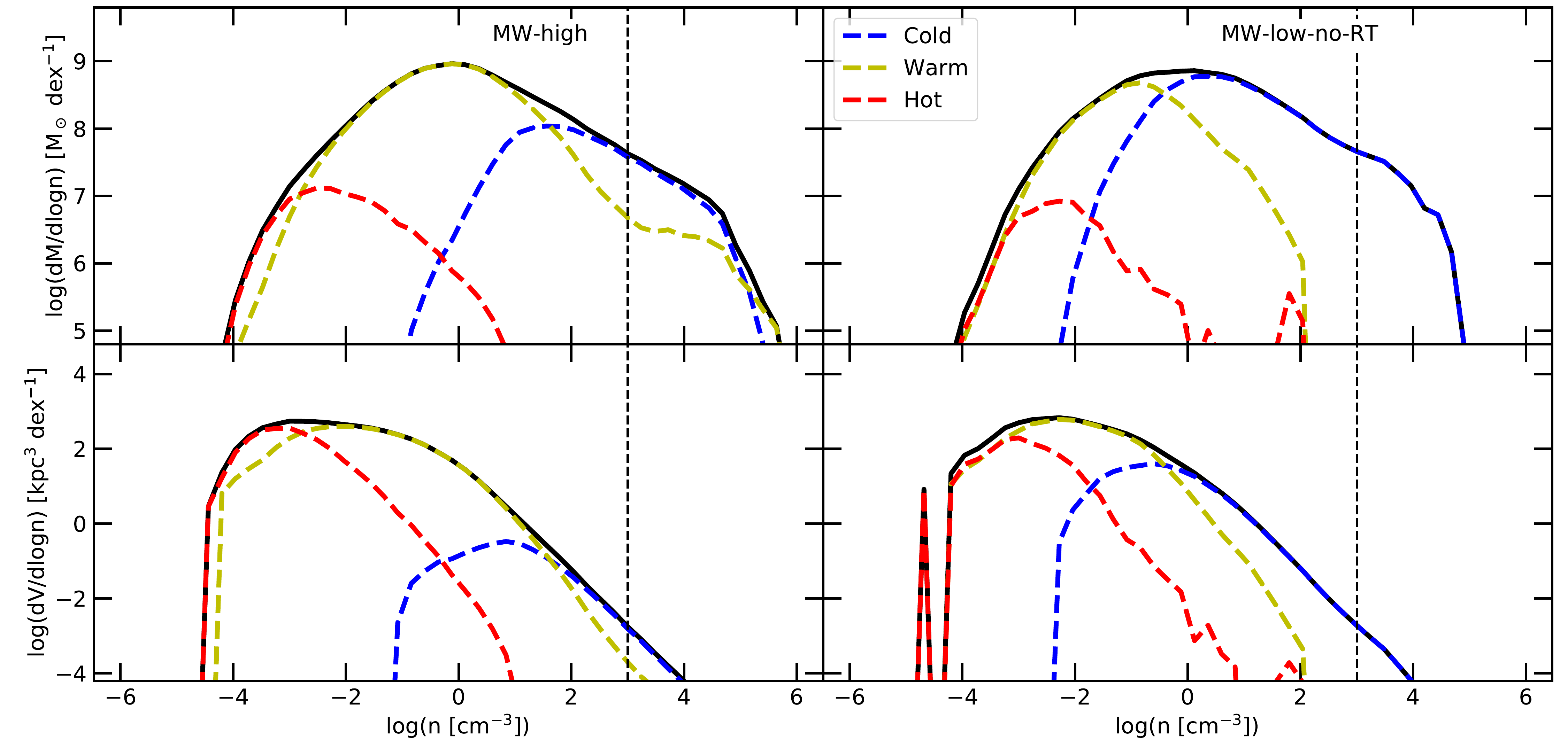}
\caption{The mass (top panels) and volume (bottom panels) distribution of gas in the disc ($R<10$ kpc and $|z| < 1.5$ kpc) in the MW-high (left panels) and MW-low-no-old (right panels) simulations. The black curve shows the distribution for all the gas in the disc, while the blue, yellow and red curves denote the mass and volume distribution of the cold ($T<2 \times 10^3$ K), warm ($2 \times 10^3 \leq T < 4 \times 10^5$ K) and hot ($T \geq 4 \times 10^5$ K) gas respectively.  }
\label{fig:struct}
\end{figure*}

\section{Results}
\label{sec:results}
We start with a visual inspection of the simulated galaxy. Figure~\ref{fig:rgb} shows the face-on false colour Red-Yellow-Blue (RYB) image (bottom panel) of the disc in the MW-high run after $1$ Gyr of evolution. The RYB image is constructed from the IR (top left panel), optical (top middle panel), and the extreme UV (EUV1, top right panel) radiation fields that are self-consistently generated in the simulation. The dust lanes are clearly visible in the optical image. Moreover, the IR and Optical images show spatial anti-correlation because dust absorbs the optical light and re-emits it in IR. The EUV1 map shows a clustered appearance with the high energy radiation escaping through the low density channels that have been cleared out by early stellar radiative and SN feedback mechanisms. Moreover, we can clearly identify both obscured and unobscured star formation by focusing on the peaks of the IR and EUV1 maps respectively.  

Figure~\ref{fig:SFR} shows the star formation rate history as a function of simulation time for the 'MW-high' (red curve), 'MW-low' (blue curve), 'MW-low-no-old' (orange curve), and 'MW-low-no-RT' (green curve) runs.  The star formation rate rises steeply in all cases within the first $10$ Myr as the disc settles. The initial starburst is about a factor of two higher in the run without radiation ($\sim 6  \ \mathrm{M}_\odot  \ \mathrm{yr}^{-1}$), compared to the run with it ($\sim 3.5 \  \mathrm{M}_\odot  \ \mathrm{yr}^{-1}$), because of the lack of photoionization and radiation pressure feedback. In these low gas surface density environments, the effect of radiation feedback is modest and it mainly affects the initial stages of the disc evolution. This is in agreement with previous studies, which have come to the same conclusion \citep{Rosdahl2015, HopkinsRT, Kannan2018}.  The runs with radiation all show a star formation rate of face-onbout $\sim 2-3  \mathrm{M}_\odot  \ \mathrm{yr}^{-1}$ throughout the simulation time. We note that the star formation rates are well converged with resolution with the MW-high simulation showing only a slight decrease in star formation compared to the corresponding low resolution run.

\begin{figure}
\includegraphics[width=\columnwidth]{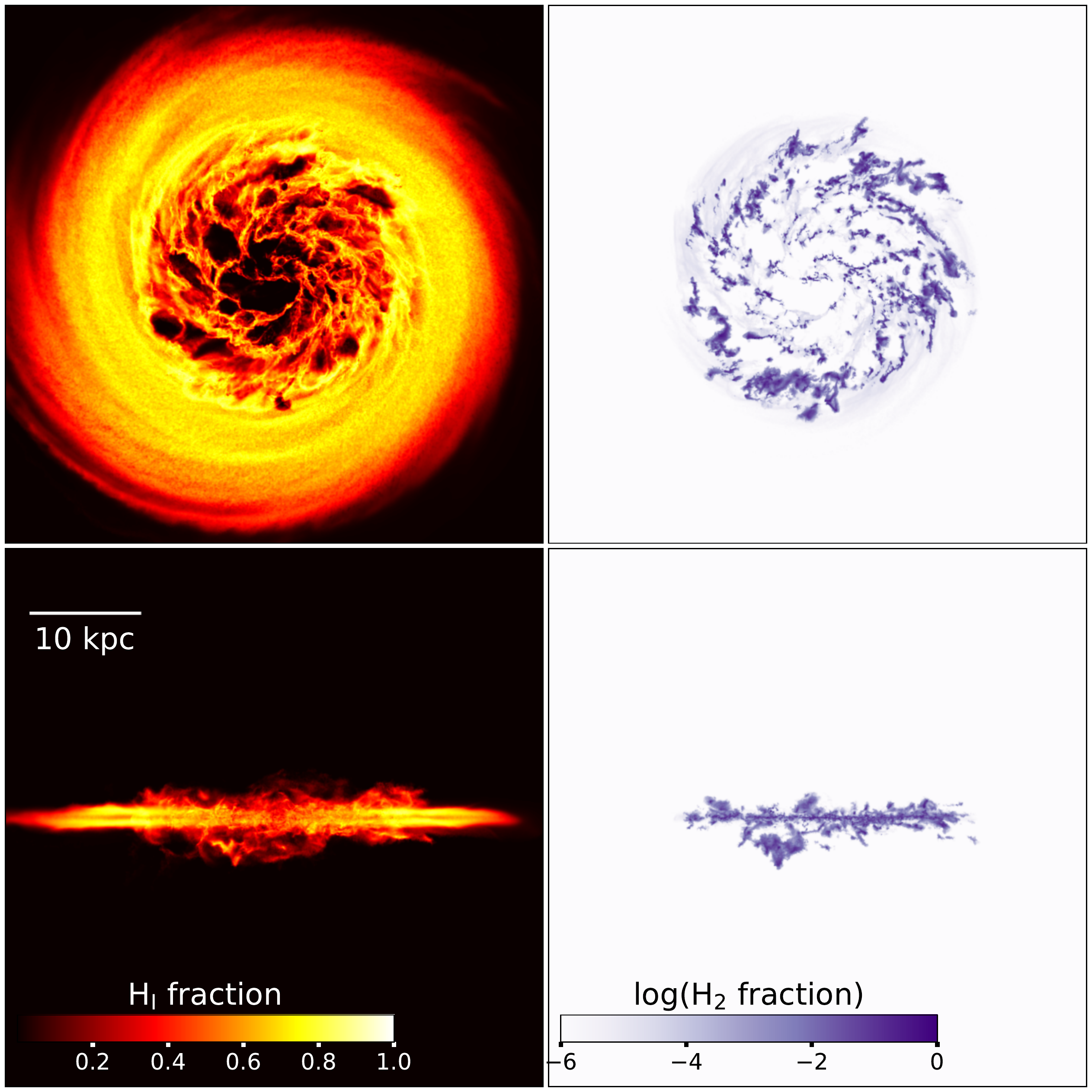}
\caption{Face-on (top panels) and edge-on (bottom panels) maps of atomic (\hi, left panels) and molecular (\HM, right panels) hydrogen fraction in the MW-high simulation.}
\label{fig:H2}
\end{figure}

\subsection{Impact of radiation fields on the structure of the ISM}

We now concentrate on the impact of radiation fields on the multi-phase structure of the ISM. To do this we compare the MW-high and the MW-low-no-RT simulations. We note that the ISM structure in the MW-low and MW-low-no-old runs are similar to the MW-high run. Accordingly, Figure~\ref{fig:phasespace} shows a two dimensional mass histogram of the gas in the temperature-density plane of the MW-high (left panel) and MW-low-no-RT (right panel) simulations. The low density gas in neither run is able to cool below $\sim 10^4$ K,  due to heating from the ultra-violet background (UVB). Only the gas above the self-shielding density threshold of $n \sim 10^{-2} \ \mathrm{cm^{-3}}$ manages to cool below this temperature.  The phase-space structure of gas above this density is markedly different in the two runs. Photoelectric heating by far-UV radiation impinging on dust grains only allows the highest density ($n \gtrsim 10^3  \ \mathrm{cm}^{-3}$) gas to cool down to $\sim 10$ K. The gas in the MW-low-no-RT simulation on the other hand reaches the minimum temperature threshold even at relatively low densities of about $n \sim 1  \ \mathrm{cm}^{-3}$. The effect of photoheating from ionizing radiation is also clearly visible in the the MW-high run. The high-density photoheated gas  around newly formed stars clusters around $\sim 2  \times 10^4$ K, driving small scale winds and thereby preprocessing the sites of SN explosions. This high-density, warm phase is nonexistent in the MW-low-no-RT run due to the lack of radiation fields.

 \begin{figure}
\includegraphics[width=\columnwidth]{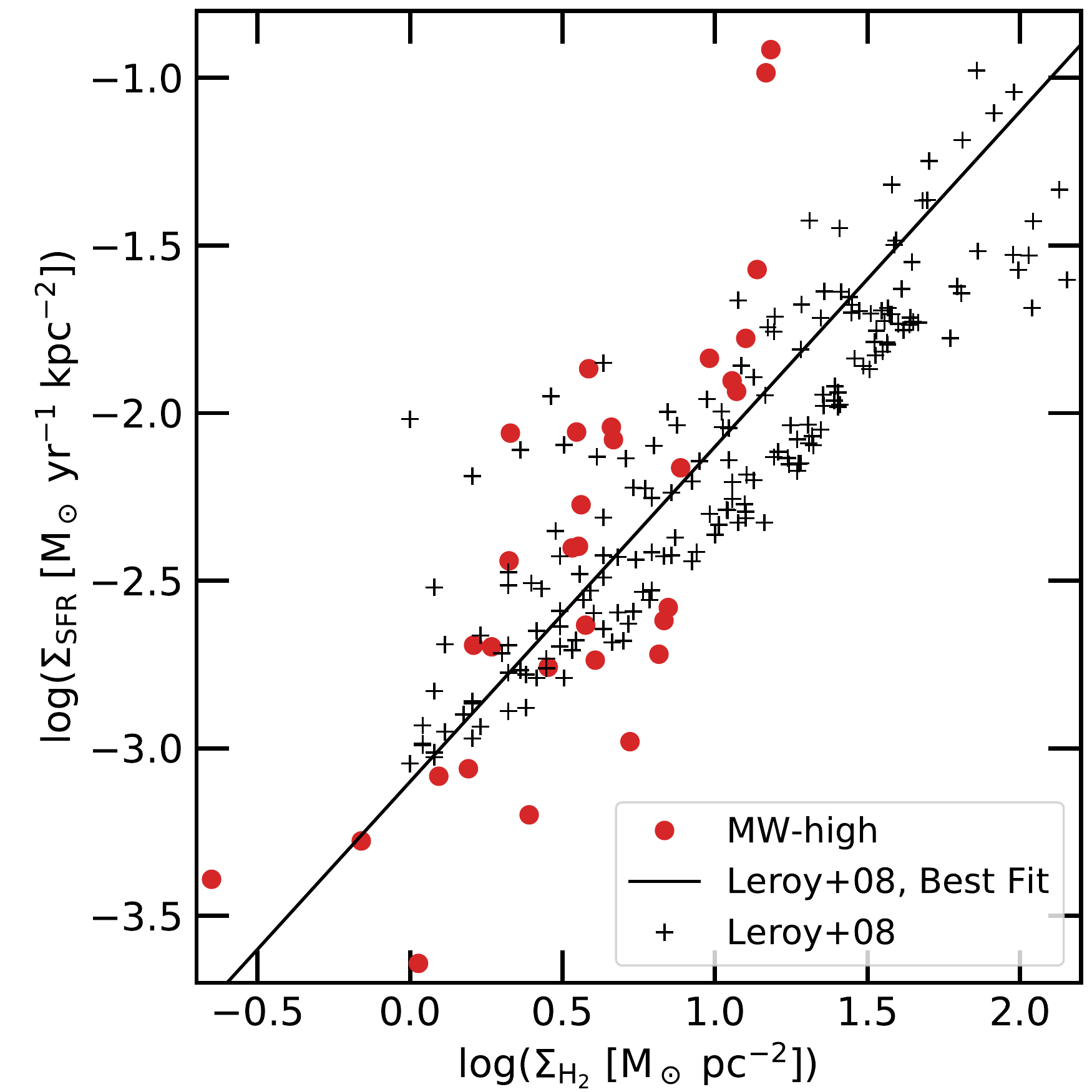}
\caption{The molecular KS relation (red points) compared to the observational estimates (black crosses and line) taken from \citet{Leroy2008}. }
\label{fig:KS}
\end{figure}

Figure~\ref{fig:struct} shows the distribution of gas mass (top panels) and volume (bottom panels) as a function of density of the gas in the disc ($R<10$ kpc and $|z| < 1.5$ kpc) at $1$ Gyr. Each distribution is further divided into the contribution from cold (blue dashed curve; $T<2 \times 10^3$ K), warm (yellow dashed curve; $2 \times 10^3 \leq T < 4 \times 10^5$ K), and hot  (red dashed curve; $T \geq 4 \times 10^5$ K) phases of the gas. The vertical dashed line denotes the star formation density threshold of $n = 10^3 \ \mathrm{cm}^{-3}$. The overall mass and volume distribution is quite similar in both the MW-high (left panels) and the MW-low-no-old runs (right panels). The distribution of the various temperature phases on the other hand is quite different, especially the cold and the warm phases. The cold phase in the run without radiation dominates the mass and volume distributions down to very low densities ($n \gtrsim 1 \ \mathrm{cm}^{-3}$), while it is dominant only up to $n \sim 100 \ \mathrm{cm}^{-3}$ in the fiducial run. This translates to cold gas mass fractions of $\sim 65 \%$ and $\sim 10 \%$ respectively.

 The warm phase follows a log normal distribution extending from about $10^{-4} \ \mathrm{cm}^{-3}$ to $10^{2} \ \mathrm{cm}^{-3}$ in the MW-low-no-RT run. However, the warm phase in the fiducial run deviates from the log-normal distribution above the star formation density threshold, with the warm gas fraction increasing at these high densities. This is due to photoheating of gas by newly formed stars. In fact at extremely high densities, where star formation is most efficient, the mass in the warm phase dominates over the cold phase, demonstrating the efficiency of photoheating. The mass fraction of the warm phase is about $35 \%$ and $90 \%$ in the MW-low-no-RT and MW-high runs respectively. The amount of gas in the hot phase is sub-dominant in both runs with less than $1\%$ mass in this phase.  It is therefore, quite clear that local radiation from stars plays an important role in governing the phase-space structure of the the ISM, mainly via photo- and photelectric- heating.

\begin{figure*}
\includegraphics[width=\textwidth]{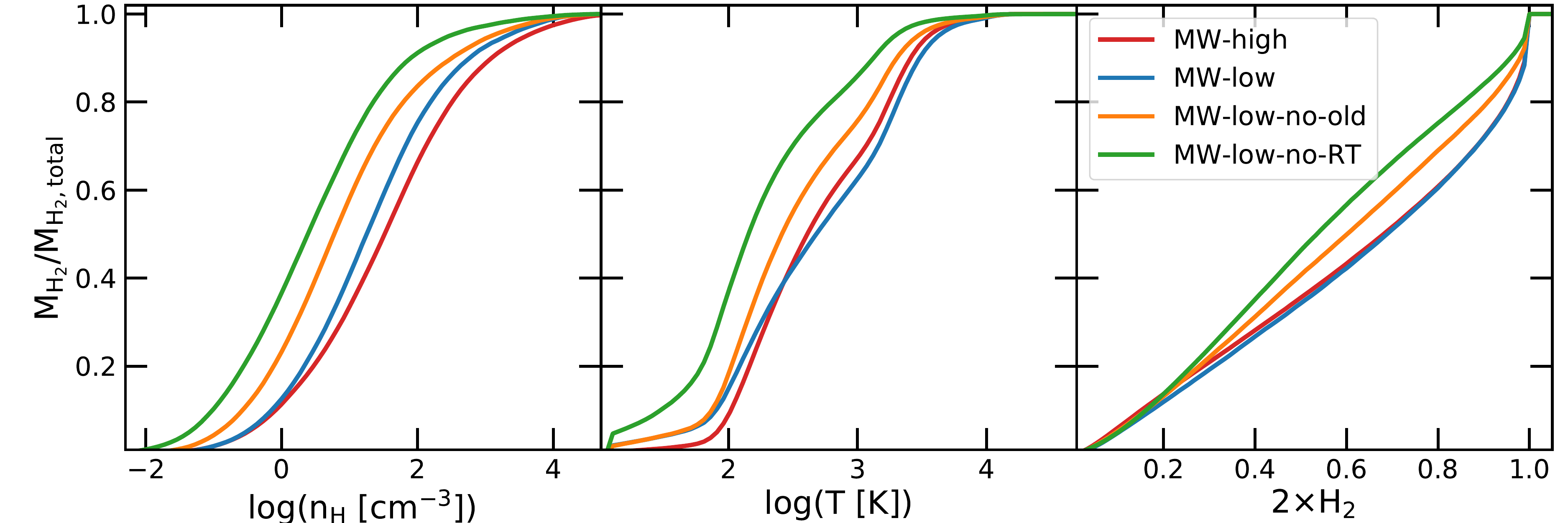}
\caption{Cumulative mass distribution of \HM~as a function of density (left panel), temperature (middle panel), and \HM~fraction (right panel) of the gas in the MW-high (red curves), MW-low (blue curves), MW-low-no-old (orange curves), and MW-low-no-RT (green curves) simulations.}
\label{fig:H2a}
\end{figure*}

\subsection{Distribution of molecular hydrogen}

 We now turn our attention to the distribution of molecular hydrogen as predicted by the non-equilibrium chemical network. Figure~\ref{fig:H2} shows face-on (top panels) and edge-on (bottom panels) maps of atomic (left panels) and molecular molecular (right panels) hydrogen fractions in the MW-high simulation at $1$ Gyr. The \hi~distribution extends to about $15$ kpc from the center of the disc, while the \HM~is only found within the central $\sim 10$ kpc. Molecular gas is also more clumpy and found in the highest density regions of the disc. The higher gas pressure in the disc midplane allows for a greater prevalence of \HM~, which can clearly be seen in the edge-on view of the disc.

A more quantitative picture can be obtained by studying the correlation between the sites of star formation and the distribution of \HM. It has been well know for some time that these quantities correlate with each other on sub-kpc  scales \citep{Bigiel2008, Leroy2008}. Figure~\ref{fig:KS} shows the molecular Kennicutt-Schmidt relation \citep{Schmidt1959, Kennicutt1989}, derived from the MW-high simulation (red points) at $1$ Gyr. For comparison, the observational estimates from \citet{Leroy2008} are shown by black crosses, with the solid black line denoting the mean relation. In order to make an accurate comparison with the observational data, we create a map of the star formation rate surface density (stars formed less $\sim 50$ Myr ago). We then pick out the peaks of this distribution\footnote{We use astrodendro, a Python package to compute dendrograms of Astronomical data (http://www.dendrograms.org/).} and compute the star formation rate and \HM~mass within a circle of radius $750$ pc around the peaks. The simulations do a very good job of matching both the scatter and the mean observational relation.

We note that the simulation overshoots the observed relation at very high star formation rate surface densities ($ \sim 0.1 \ \mathrm{M}_\odot \ \mathrm{yr}^{-1} \ \mathrm{kpc}^{-2}$), corresponding to gas depletion times ($\tau_\mathrm{dep}$) of $\sim 0.2$ Gyr, compared to $\tau_\mathrm{dep} \sim 2$ Gyr in the other regions of the galaxy. These points lie in the center of the disc where the gravitational potential is the highest. This increases the
molecular gas pressure, requiring a higher star formation rate per unit molecular
mass to offset enhanced turbulent dissipation and cooling \citep{Ostriker2011}. There is observational evidence for the breakdown in this relation in galaxies above the star-forming main sequence \citep[e.g., ][]{Genzel2010, Genzel2015} and in galactic centers \citep{Jogee2005, Leroy2013, Utomo2017}. The $\Sigma_\mathrm{SFR} \sim 0.1 \ \mathrm{M}_\odot \ \mathrm{yr}^{-1} \ \mathrm{kpc}^{-2}$ is at the low end of the distribution found in local star bursting galaxies \citep{ulirgs2016}, where the local gas depletion time is found to be about $\sim 0.2$ Gyr, in agreement with our results.

Finally, we show in Figure~\ref{fig:H2a} the cumulative mass distribution of \HM~as a function of density (left panel), temperature (middle panel), and \HM~fraction (right panel) of the gas in the MW-high (red curves), MW-low (blue curves), MW-low-no-old (orange curves), and MW-low-no-RT (green curves) simulations. The various physical models exhibit rather different \HM~distributions, while the fiducial simulation shows good convergence as the resolution is increased.   Without any radiation fields, \HM~begins to form at relatively low densities of $n_\mathrm{H} \sim 10^{-2} \ \mathrm{cm}^{-3}$, which is the self-shielding density to UVB, while the runs with RT delay the formation for \HM~to above $n_\mathrm{H} \sim 0.1 \ \mathrm{cm}^{-3}$. The higher the radiation intensity, the larger the density of the gas needs to be in order to self-shield, shifting the \HM~mass distribution.  The temperature distribution is quite intriguing, with the runs with  radiation fields shifting the \HM~mass distributions to higher temperatures.  This can be explained by the fact that stars are generally formed in the high-density gas implying that the radiation field strengths are higher in this part of the phase space. The high-energy \HM~and \hi~ionizing photons heat up the gas to temperatures of $\sim 10^4$ K, but they also have the highest interaction cross- sections, implying that they are attenuated close to the source. The far UV photons on the other hand have larger mean free paths as they are attenuated only by dust grains, allowing them to penetrate the self-shielded high-density cold gas. They heat up the gas through photoelectric heating (without affecting the ionization state of the gas), shifting the \HM~mass distribution to higher temperatures. Finally, radiation also shifts the mass distribution to higher \HM~number densities because this translates to larger column densities, which in turn means more self-shielding from \HM~dissociating/ionizing photons.


\subsection{Dust distribution in the ISM}

\begin{figure}
\includegraphics[width=\columnwidth]{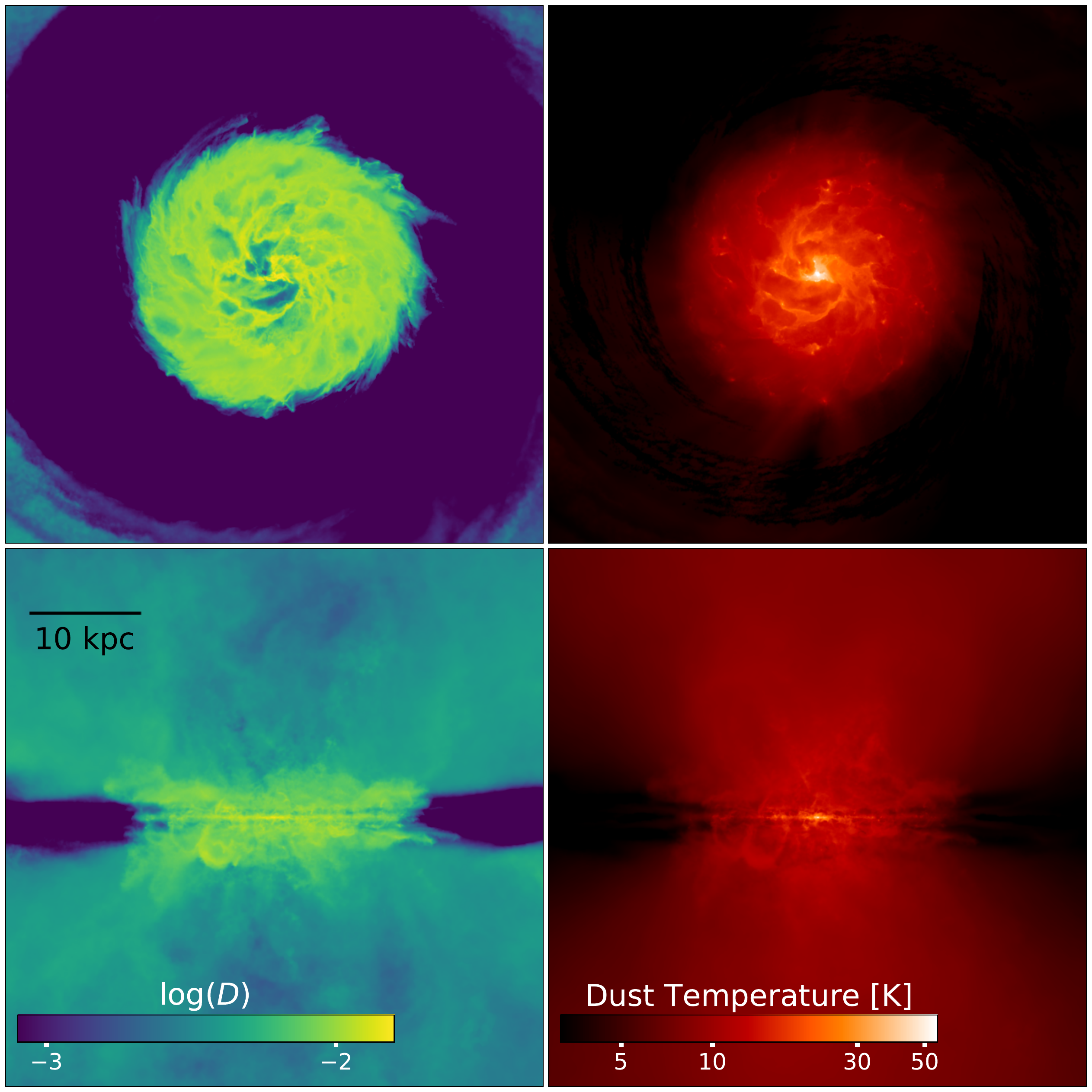}
\caption{Face-on (top panels) and edge-on (bottom panels) maps of the dust to gas ratio (left panels) and dust temperatures (right panels) in the MW-high simulation.}
\label{fig:dustMap}
\end{figure}

We now consider the properties of the dust distribution in the ISM. Figure~\ref{fig:dustMap} shows face-on (top panels) and edge-on (bottom panels) maps of the dust-to-gas ratio ($D$; left panels) and the dust temperature ($T_d$, right panels) in the MW-high simulation at $1$ Gyr. The dust-to-gas ratio hovers around a value of $\gtrsim 0.01$ over the entire inner star forming disc ($\leq 10 $ kpc). The outer low density, warm \hi~disc (n$_\mathrm{H} \sim 1 \, \mathrm{cm}^{-3} $; $\mathrm{T} \, \sim 10^4$~K) is devoid of dust, because there is no star formation activity in this region. Interestingly, dust is also present outside the disc, in the CGM of the galaxy. This dust has formed in the disc and was blown out by stellar feedback driven outflows.  A similar trend is also found in the dust temperature maps. The radiation fields from stars from the inner disc heats up the dust to about $\sim 20$~K on average. The central star forming region shows much higher dust temperatures of $\sim 30-40$~K. The edge-on view shows that the temperature of extraplanar dust is higher, perpendicular to the disc and colder along it. This is because the radiation from stars preferentially escapes along the angular momentum axis of the disc, due to the lower optical depth in this direction.

\begin{figure}
\includegraphics[width=\columnwidth]{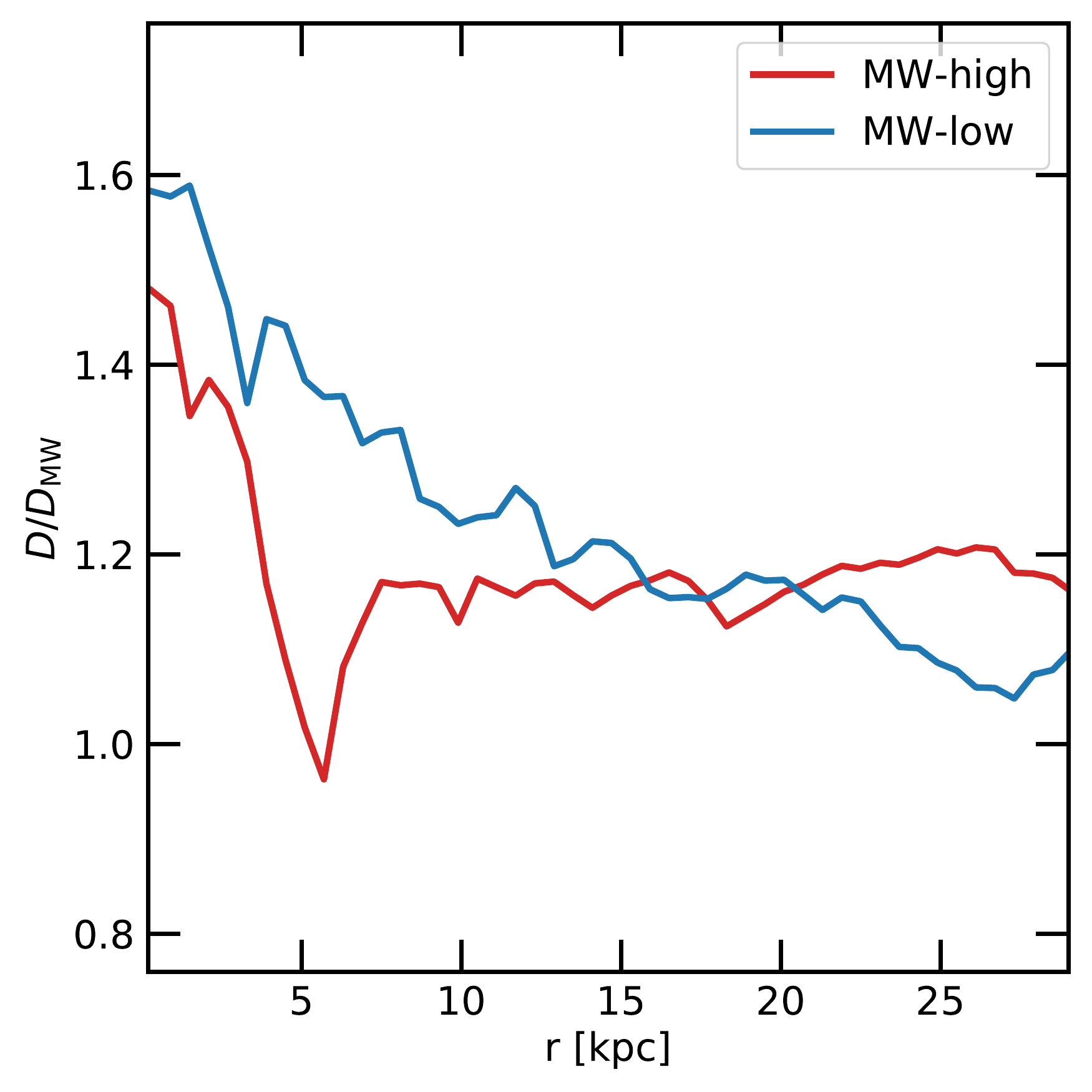}
\caption{Dust to gas ratio relative to the canonical MW value (0.01) as a function of radial distance in the MW-high (red curve) and MW-low (blue curve) simulations. }
\label{fig:DGR}
\end{figure}

Figure~\ref{fig:DGR} shows the dust mass weighted radial profile of  $D$ relative to the canonical Milky-Way value ($D_\mathrm{MW}  = 0.01$) in the MW-high (red curve) and MW-low (blue curve) simulations at $1$ Gyr. Both simulations show an increase in $D$ towards the center of the galaxy, where the star formation rate is higher, reaching values of up to  $D \sim 0.015$. The ratio slightly decreases as we get to larger galactocentric radius to a value of about $\sim 1.2$ at a distance of $25$~kpc. The radial dependence of $D$ in our simulations is shallower than what is observed in the MW \citep{Gianetti2017}. This is due to the fact that we do not have pristine circum-galactic gas around the disc, which would cool and mix with the gas in the disc, diluting its metallicity.  The various resolutions give slightly different results, with the low resolution run showing a larger radial dependence than the MW-high simulation. This indicates that the dust produced in the central star forming region is more efficiently transported outwards in the high resolution simulation compared to the low resolution run.

\begin{figure*}
\includegraphics[width=\textwidth]{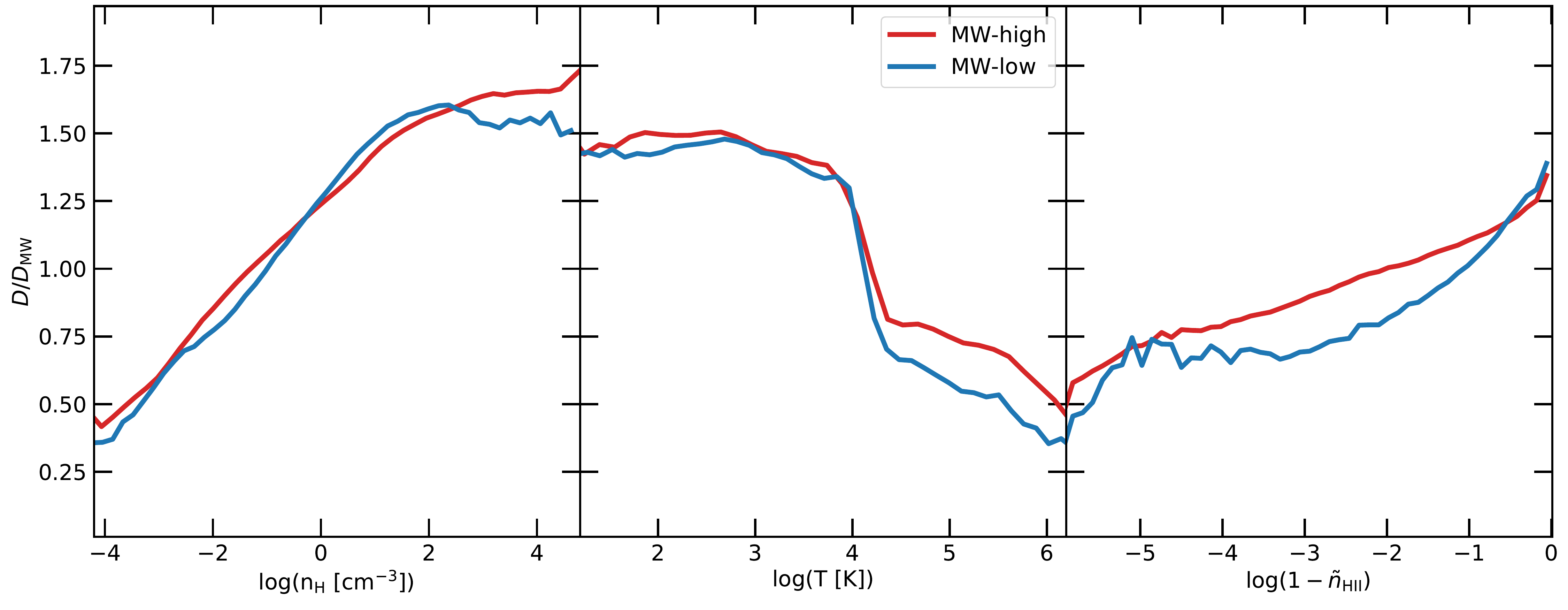}
\caption{Dust to gas ratio relative to the canonical MW value as a function of density (left panel), temperature (middle panel), and ionization state (right panel) of the gas in the MW-high (red curves) and MW-low (blue curves) simulation.}
\label{fig:dgrTrho}
\end{figure*}

Figure~\ref{fig:dgrTrho} plots $D/D_\mathrm{MW}$ as a function of the gas density (left panel), gas temperature (middle panel) and the ionization state of the gas (right panel) in the low and high resolution fiducial simulations. The results are taken after $1$~Gyr of simulation time and considers only gas present in the disc, which is defined by $R<10$~kpc and $-1.5< z (\mathrm{kpc}) < 1.5$. $D$ is a strong function of the density of the gas,  with the highest density gas, $\mathrm{n_H} \gtrsim 10^4 \ \mathrm{cm}^{-3}$, having a value $D\sim 0.016$. The value decreases as the density of the gas decreases with  $D\lesssim 5 \times 10^{-3}$ at  $\mathrm{n_H} \lesssim 10^{-4} \ \mathrm{cm}^{-3}$.  The dust distribution shows a very interesting dependence on the temperature of the gas.  Below $T\lesssim 10^4$~K, $D$ is constant with a value of $\sim 0.014$. There is a precipitous drop by almost a factor of two ($D \sim 7 \times 10^{-3}$) at $10^4$~K.  It is noteworthy that this transition happens at roughly the photoheating temperature of soft spectral sources like high mass stars and is also close to the collisional ionization temperature of atomic hydrogen at the typical densities in the ISM. In essence, this is the temperature at which the ISM transitions from a  largely neutral state to an ionized one. This clearly indicates that the dust abundance correlates with the ionization state of the gas, which is shown explicitly in the right panel of Figure~\ref{fig:dgrTrho}.  $D$ reduces from $\sim 0.014$ in neutral gas to $\sim 5 \times 10^{-3}$ in the highly ionised medium. The only processes capable of heating the gas to these temperatures are photoheating and SN explosions, which are spatially correlated. The destruction of dust in SN shocks (as given by Eq.~\ref{eq:SNdust}) and dust sputtering at high temperatures (Eq:~\ref{eq:dustsputter}) reduce the dust content around sites of star formation. We note that dust destruction seems to be slightly more effective in the low resolution runs compared to the high resolution one, as evidenced by the higher $D$ in the high temperature, high ionization gas in the MW-high simulation. These plots clearly show that $D$ is not constant, as has been assumed in many previous simulation works \citep[e.g., ][]{Rosdahl2015, Costa2018}. While a metal dependent $D$ might be more physical \citep{Bieri2017}, it will not reproduce the complex dust distributions we see in our simulations.  The dust physics must be modeled explicitly in order to obtain a correct picture of dust and radiation-dust interactions in galaxies \citep{Barnes2018}.

\begin{figure*}
\includegraphics[width=\textwidth]{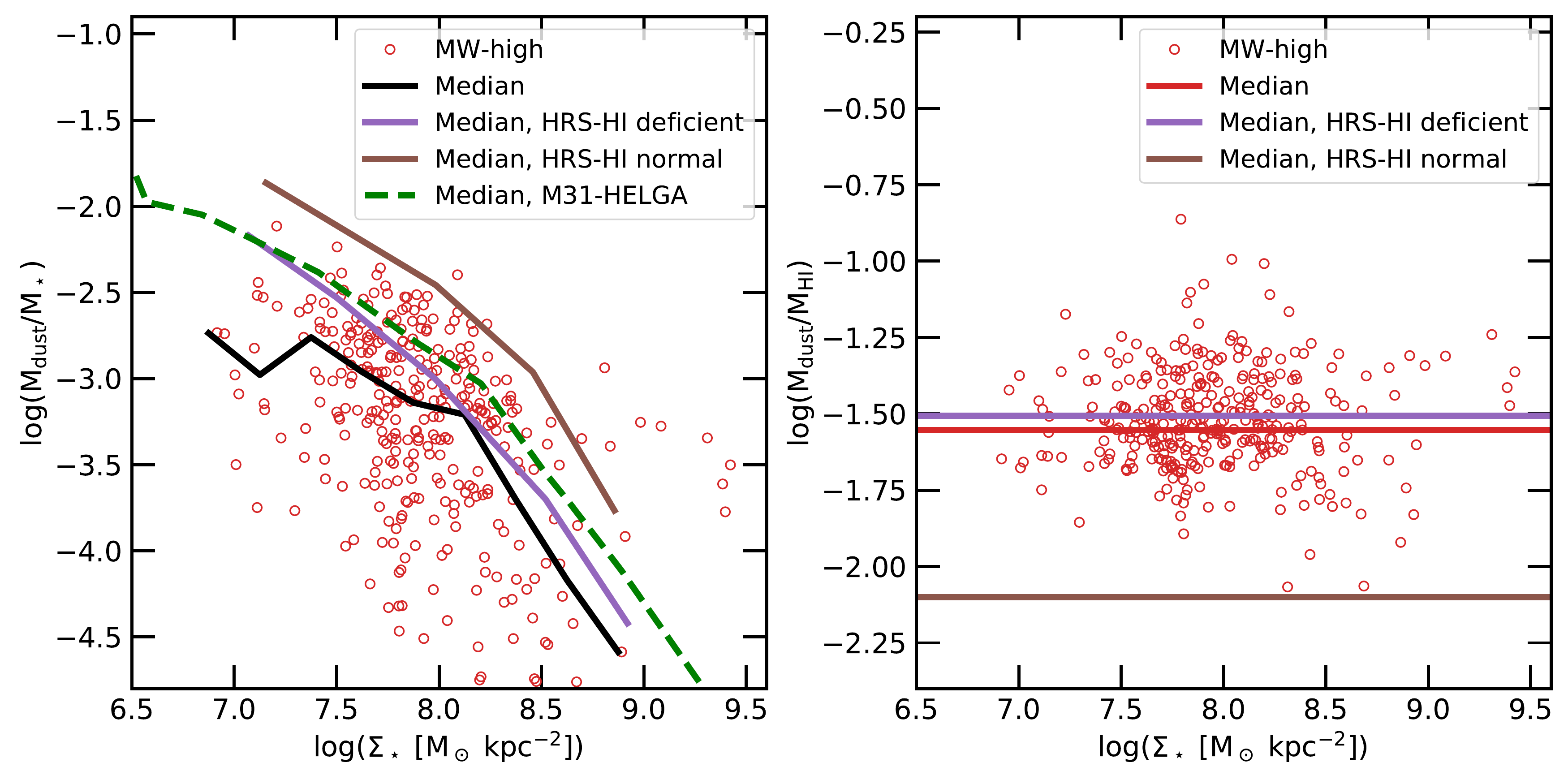}
\caption{Ratio of the dust to stellar mass (left panel) as a function of the stellar surface density. The red open circles show the results from the MW-high simulation.  The thin black line shows the median of the simulation results, the purple and brown curves show the median relation for the \ion{H}{I} deficient and \ion{H}{I} normal galaxies from the HRS sample \citep{Cortese2012}, and the dashed green line denotes the median relation for M31 taken from the HELGA sample \citep{Viaene2014}. The right panel shows the ratio of the dust to atomic hydrogen mass in the galaxy as a function of stellar surface density. The simulated median value of this ratio (red solid curve) is close to the observational estimates for the \ion{H}{I} deficient sample from HRS (purple curve).}
\label{fig:Mdust}
\end{figure*}

\begin{figure*}
\includegraphics[width=\columnwidth]{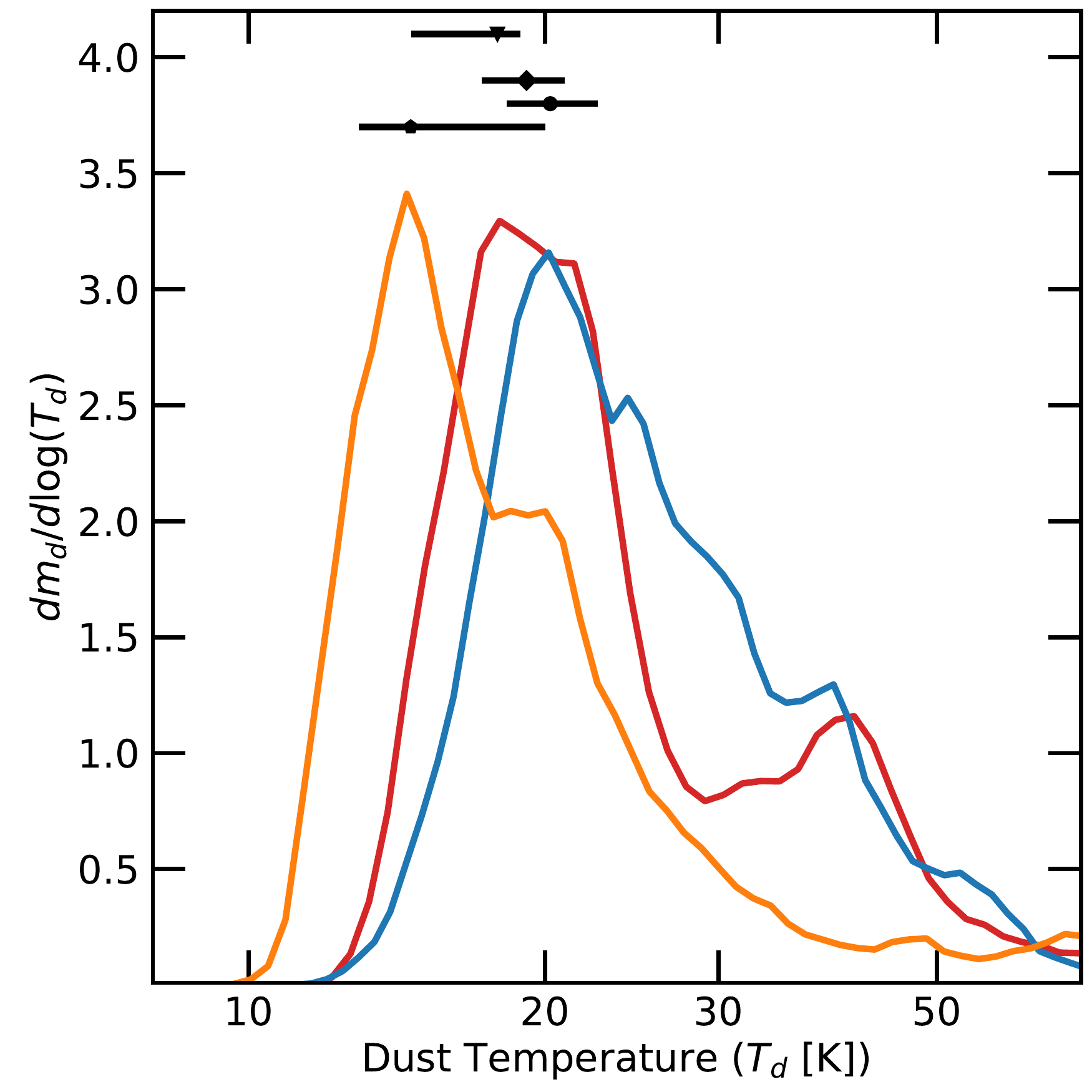}
\includegraphics[width=\columnwidth]{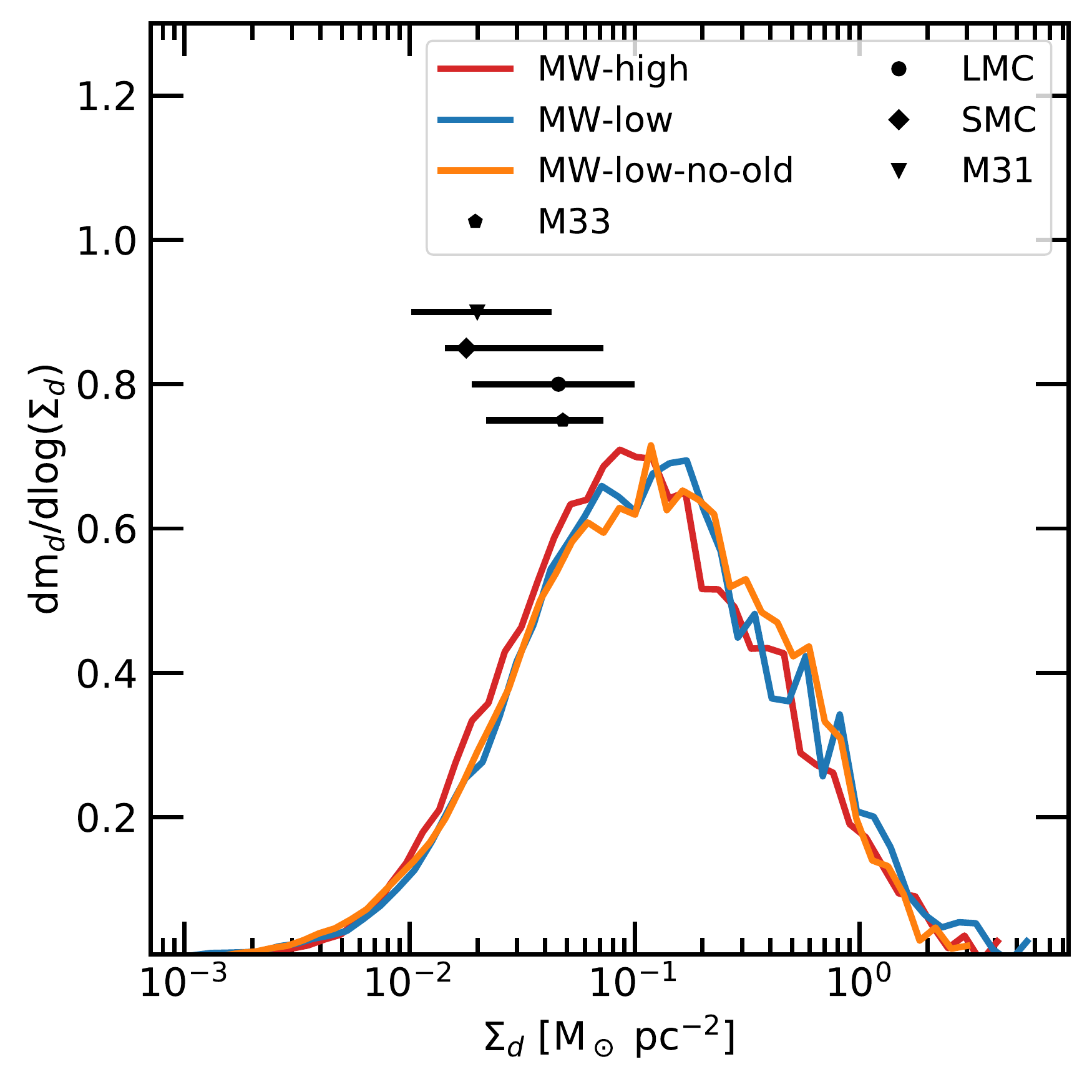}
\caption{Normalised dust temperature (left panel) and surface density (right panel) distributions in the MW-high (red curve), MW-low (blue curve), and MW-low-no-old (orange curve) simulations. The black points show the observational estimates for the LMC (circle), the SMC (diamond), M33 (pentagon), and M31 (inverted triangle) taken from \citet{Utomo2019}. }
\label{fig:histdust}
\end{figure*}

We now turn our attention to the resolved scaling relations of dust with respect to the stellar and gas mass of the galaxy. Figure~\ref{fig:Mdust} shows the ratio of the dust to stellar mass (left panel) as a function of the stellar surface density for the MW-high simulation (red open circles) at $1$ Gyr. This plot is generated by dividing the inner disc ($R<10$~kpc) into $1 \, \mathrm{kpc}^2$ areas and we calculate the quantities for each portion. The thin black line shows the median of the results from our simulation. We compare our results to the observational estimates taken from the {\it Herschel} Reference Survey \citep[{\it HRS};][]{Boselli2010, Cortese2012}  and The {\it Herschel} Exploitation of Local Galaxy Andromeda project \citep[{\it HELGA};][]{Fritz2012, Viaene2014}. The {\it HRS} sample consists of $\sim 300$ nearby galaxies and is sensitive to the cold dust component. The {\it HELGA} project focuses on the characteristics of the extended dust emission of Andromeda at a high spatial resolution ($\sim 0.6$~kpc).  These datasets therefore cover both the galaxy integrated and resolved properties of dust in the nearby Universe.  The simulations show that the dust-to-stellar mass ratio decreases with increasing stellar surface density and is in rough agreement with both the {\it HELGA} (dashed green curve) and the \ion{H}{I}-normal (brown curve) and \ion{H}{I}-deficient (purple curve) {\it HRS} observational samples, although the median relation is slightly lower than what is observed.

The right panel of Figure~\ref{fig:Mdust} plots the ratio of the dust to atomic hydrogen mass as a function of the stellar mass surface density. This ratio is roughly constant at a value of about $\sim 10^{-1.55}$ (red solid curve), irrespective of the stellar surface density. This is a bit higher than the average value for \ion{H}{I}-normal systems ($\sim 10^{-2.1}$) but has remarkable  agreement with the results for \ion{H}{I} deficient galaxies ($\sim 10^{-1.50}$) in the {\it HRS} sample \citep{Cortese2012}. The scatter in the relation is also quite similar to the observational estimates.  We note that our results are more in agreement with the \ion{H}{I}-deficient galaxies, because, the lack of pristine gas inflow from the CGM in our simulation makes the gas in the center more dust enriched than usual, mimicking the properties of galaxies that are deficient in atomic hydrogen. These two plots together, show that the dust to stellar mass ratio decreases due to a corresponding decrease in the gas fractions with increasing stellar surface density, which has been shown to be true observationally \citep{Catinella2010, Cortese2011}.

\begin{figure*}
\includegraphics[width=\textwidth]{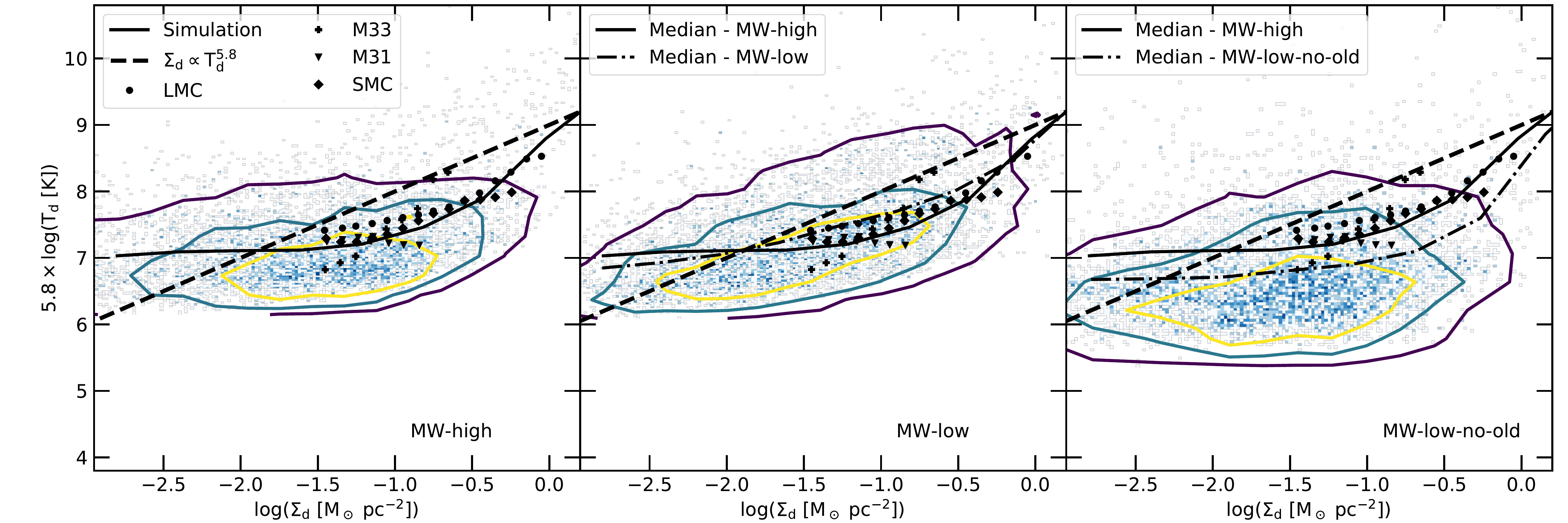}
\caption{2D histograms showing the dust temperature-dust surface density relation in the MW-high (left panel), MW-low (middle panel), and MW-low-no-old (right panel) simulations.  The contours indicate the $20\%$ (purple contour),  $50\%$ (blue contour), and $80\%$ (yellow contour) enclosed fractions of the distribution. The solid black line denotes the median of the distribution, and the dashed black curves indicates $\Sigma_d \propto T_d^{5.8}$, valid for dust heated by star formation. The black points show observational estimates for the LMC (circles), the SMC (diamonds), M33 (pentagons), and M31 (inverted triangles) taken from \citet{Utomo2019}.}
\label{fig:dT}
\end{figure*}


\begin{figure}
\includegraphics[width=\columnwidth]{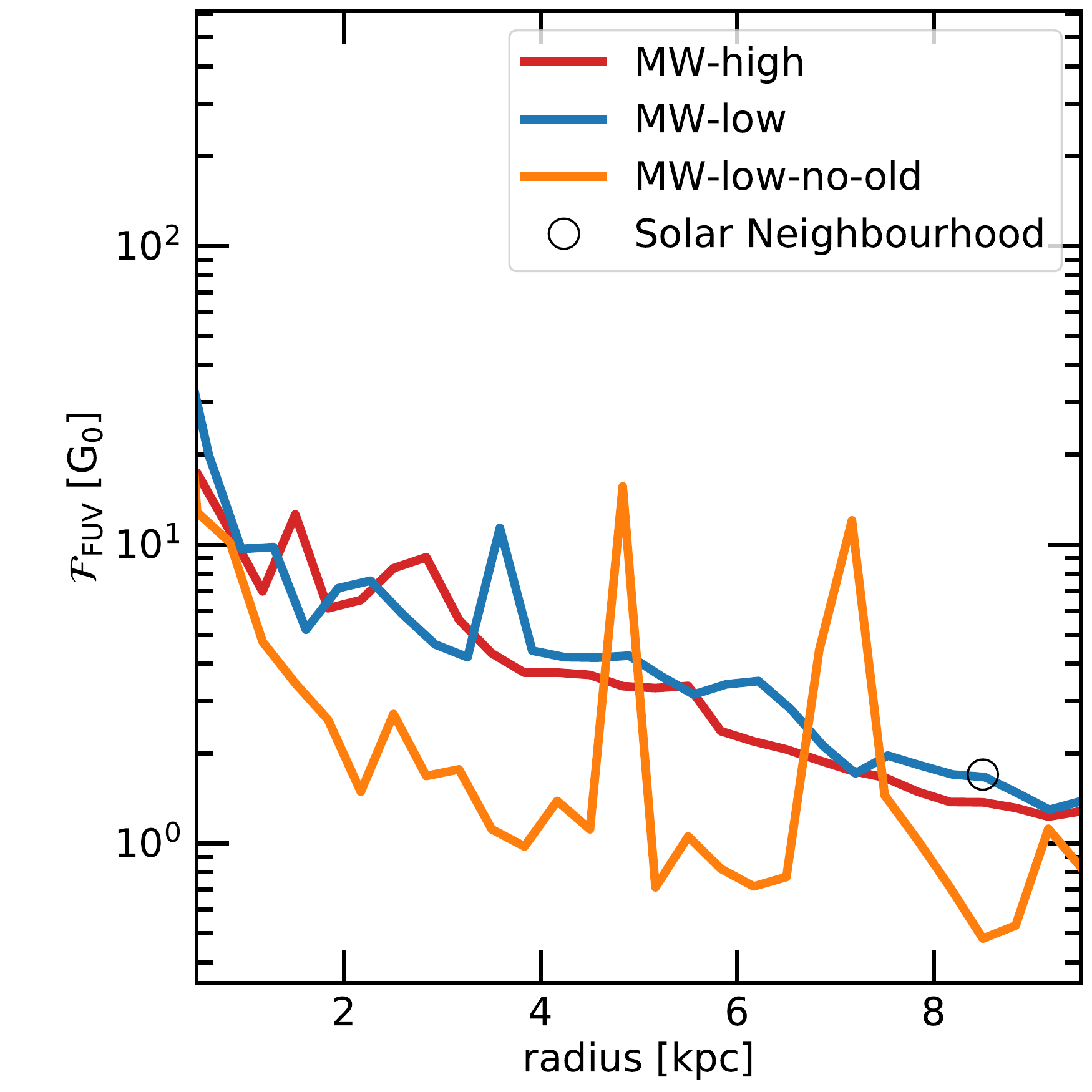}
\caption{Value of the FUV flux  (in units of G$_0$) as a function of radial distance for the MW-high (red curve) and MW-low (blue curve) simulations. The solar neighbourhood value is indicated by the black point \citep{Habing1968}.}
\label{fig:G0}
\end{figure}

\begin{figure}
\includegraphics[width=\columnwidth]{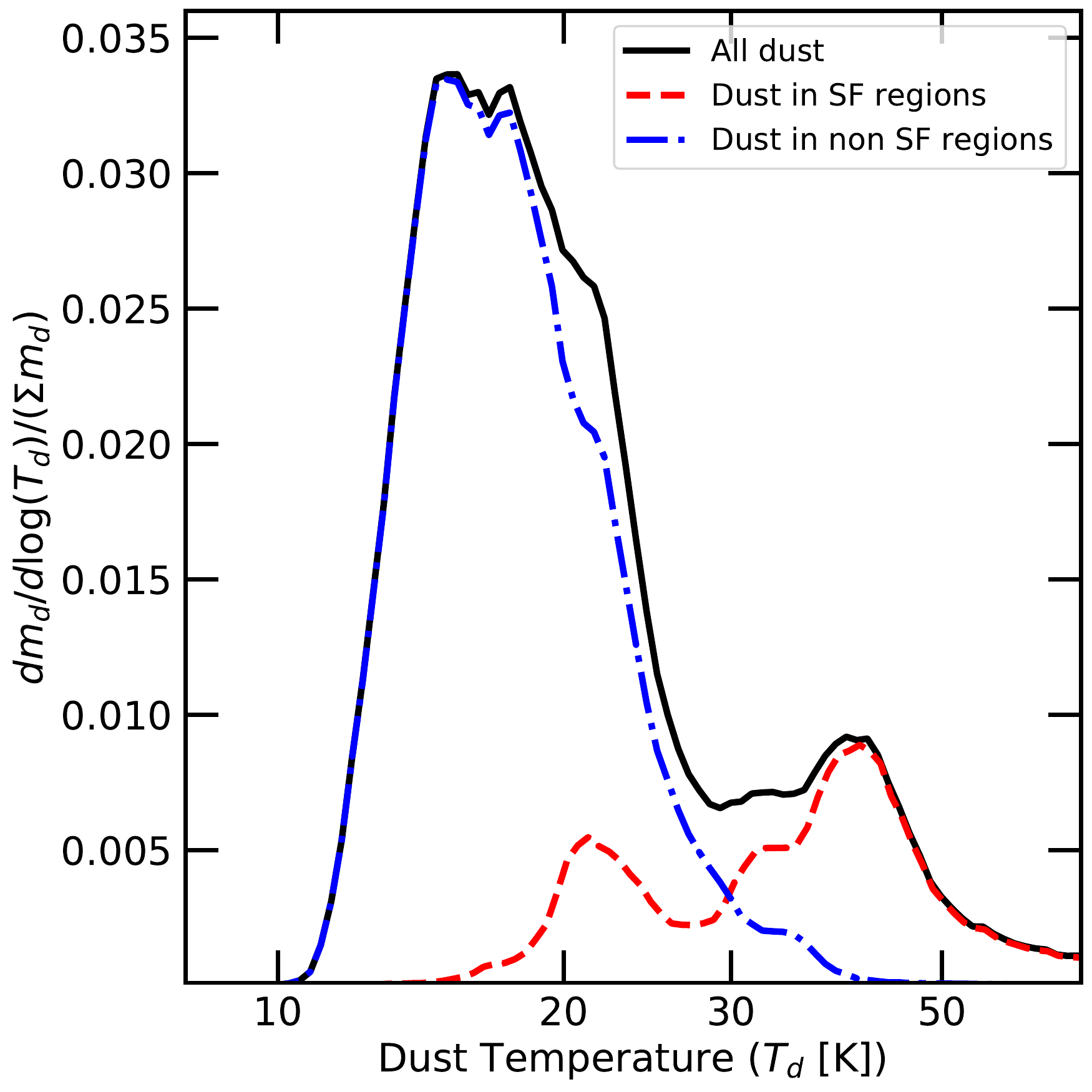}
\caption{Temperature distribution of all the dust in the MW-high simulation (black curve), split by dust temperature in star-forming (red dashed curve) and non star-forming regions (blue dot-dashed curve). }
\label{fig:dTprofile}
\end{figure}

\subsection{Dust temperature distribution and the contribution from old stellar populations}
This section examines the dust temperatures ($T_d$) predicted by our simulations. Dust temperature depends on the composition and sizes of the dust grains, and on the interstellar radiation field (ISRF).  Most of the far-UV and visible light from stars is absorbed by dust and reradiated in the IR regime. It is well known that the dust in galaxies is in radiative equilibrium with the ISRF \citep{Draine2003}. At high redshifts, various works have inferred the dust mass and temperatures from the integrated SEDs of unresolved objects \citep{Scoville2014, Genzel2015}. Resolved studies of dust properties are possible for the local group galaxies like the LMC, the SMC, \citep[e.g., ][]{Meixner2013, Chastenet2017}, M31 \citep[e.g., ][]{Fritz2012, Draine2014}, and M33 \citep{Braine2010, Xilouris2012}.  

In Figure~\ref{fig:histdust} we show the normalised histogram of dust mass as a function of dust temperature (left panel) compared to observational estimates inferred from the {\it Herschel} telescope for the LMC (black circle), the SMC (black diamond), M31 (inverted triangle), and M33 (black pentagon)  \citep{Utomo2019}. In order to be consistent with the observations we divide the inner $10$~kpc of the disc into regions, $167 \times167 \, \mathrm{pc}^2$ in size and calculate the average dust, temperature, surface density, and mass for each portion of the disc. The histograms in both the MW-high (red curves) and MW-low (blue curves) peak at about $\sim 20$~K, with a high temperature tail extending to $T_d \gtrsim 50$~K.  The mean dust temperatures in the local group galaxies lie between $15-22$~K, which is consistent with our simulation results. The higher resolution simulation shows slightly lower dust temperatures, but the high temperature wing exists in both simulations. In order to understand the role of old stars in dust heating, we also show results from the MW-low-no-old simulation (orange curves), which as described before, only considers radiation from stars that have formed after the start of the simulation. The fiducial models meanwhile account for radiation from both the newly formed stars and from stars that were present in the initial conditions. Not considering the radiation filed from old stars clearly underpredicts the dust temperatures, with the peak at $T\lesssim 15$~K. We do note that the high temperature wing, although still present in the MW-low-no-old simulation, is less prominent than in the fiducial runs.  In the right panel of Figure~\ref{fig:histdust}, we show the corresponding normalised histogram of dust mass as a function of dust surface density in the simulations considered. The dust surface densities are a bit more converged, but are about a factor of $\sim 2$ higher than what is observed in the local group galaxies. We note that there is essentially no difference between the dust surface density distributions in the three runs. Therefore, the difference in the dust temperature distributions can be attributed to the increase in the ISRF in the fiducial runs compared to the run without radiation fields from old stars.

This is seen more clearly in Figure~\ref{fig:dT}, which shows the two dimensional histogram of the dust temperature and dust surface densities, compared to the observational estimates.  The contours indicate the $20\%$ (purple contour),  $50\%$ (blue contour), and $80\%$ (yellow contour) enclosed fractions of the distribution. The solid black line denotes the median of the distribution, and the dashed black curve is the $\Sigma_d \propto T_d^{5.8}$ relation, valid for dust heated by star formation \citep{Utomo2019}. The fiducial high resolution simulation (MW-high; left panel) shows a reasonable match with the observational results. The MW-low simulation (middle panel) shows similar results, with the median relation being quite close to the high resolution simulation. This demonstrates that the dust temperature and surface densities are well relatively converged with respect to resolution. The run without radiation fields from old stars (MW-low-no-old; right panel), on the other hand, shows consistently lower dust temperatures by about $\sim 0.5$~dex in the entire dust surface density range. The difference is larger at lower surface densities implying that the contribution to dust heating from old stars is more important in low dust surface density regions. Therefore, we conclude that old stellar populations play a non-negligible role in heating up the dust in galaxies \citep{Groves2012, Kirkpatrick2014, Leja2019}.

This difference in the dust temperatures is caused by the difference in the ISRFs in the simulations with and without radiation contribution from old stars as seen in Figure~\ref{fig:G0}, which plots the flux of the far-UV radiation field ($5.8-11.2 \, \mathrm{eV}$) in units of the Habing value \citep[$G_0 = 1.6 \times 10^{-3} \, \mathrm{erg} \, \mathrm{s}^{-1} \, \mathrm{cm}^{-2}$; ][]{Habing1968} as a function of galacto-centric distance. Both the fiducial simulations (MW-high: red curve and MW-low: blue curve) show a value of the ISRF that is about a factor of $5-10$ higher than the MW-low-no-old (orange curve). This translates to an increase in the temperature of the dust by a factor of about $\sim 1.5-2$, which explains the differences in Figs.~\ref{fig:histdust}~and~\ref{fig:dT}. We also note that the the fiducial runs match the solar neighbourhood value of the ISRF (black circle), while the MW-low-no-old run does not. 


Finally, Figure~\ref{fig:dTprofile} shows the mass weighted dust temperature distribution normalised to the total dust mass (solid black curve), similar to Figure~\ref{fig:histdust}, but now decomposed into dust in star-forming (dashed red curve) and non star-forming (blues dot-dashed curve) regions. Star-forming regions are taken to be any dust present within $1$~kpc of the sites of star formation. These sites are determined by  creating a map of the star formation rate surface density (stars formed less $\sim 50$ Myr ago). The peaks of this distribution are taken as centers of the sites of star formation.  As expected, the star-forming regions tend to have warmer dust temperatures of about $\sim 40$~K, due to the higher radiation field strengths, while the old stellar population heats the rest of the dust to about $\sim 18$~K. Since most of the disc is in non star-forming regions, the dust temperature distribution peaks at this temperature.\footnote{We note that the secondary peak in the dust temperature distribution at about $\sim 20$~K in the star forming regions is because the choice of $1$~kpc radius around star formation peaks is quite arbitrary, implying that the selected regions will also be polluted by dust heated by the old stellar population.}  This result is in agreement with observational estimates of dust temperature in star-forming regions in local group galaxies \citep{Relano2009, Taba2010, Utomo2019}.  This explains the cold dust temperature peak and the warm dust temperature wing seen in the dust temperature distribution in our simulations.

\section{Summary and Conclusions}
\label{sec:conc}
In this paper we present a state-of-the-art model to self-consistently treat the effects of radiation fields, dust physics, and molecular chemistry in the interstellar medium of galaxies. This work builds on the  the resolved ISM model {\it SMUGGLE}, introduced in \citet{smuggle}. We retain the star formation,  SN and stellar wind feedback prescriptions, and the metal enrichment strategies of {\it SMUGGLE}. We replace the sub-grid radiation feedback prescriptions for photoheating and radiation pressure with accurate radiation hydrodynamics.  Our model also introduces more realistic prescriptions for the gas cooling and heating processes that occur in the low temperature ($\lesssim 10^4$~K) ISM by replacing {\sc cloudy} fits to the cooling rates with a model for cooling via fine structure metal lines combined with cooling from molecular hydrogen and dust-gas collisions. Photo-heating from ionizing radiation and photoelectric heating from far-UV radiation impinging on dust grains are calculated from the radiation fields  generated from the stars in the galaxy. Moreover, as the abundance of molecular hydrogen and dust are important for determining the cooling and heating rates of gas in the ISM, we endeavor to model them in a self-consistent manner.   The abundance of molecular hydrogen is estimated using a non-equilibrium thermochemical network. The ionization and heating rates are set by the local radiation field strength of each cell. Dust is modeled using the method outlined in \citet{McKinnon2016}, which accounts for dust production in SN and AGB stars, dust destruction via SN events, and thermal sputtering in the high temperature gas. In addition, we introduce a scheme to estimate the dust temperature distribution. The dust abundances and temperatures in turn control the formation rate of \ion{H}{$_2$} and the cooling rate via dust-gas collisions. 

We tested this model in simulations of an isolated non-cosmological Milky Way-like disc. Our main findings are:
\begin{enumerate}
\item Star formation rates are maintained at a value of about $\sim 2-3 \ \mathrm{M}_\odot \ \mathrm{yr}^{-1}$ throughout the entire simulation time, consistent with observational data.  The initial starburst is about a factor of $2$ higher in the run without radiation fields compared to the runs with it, because of the lack of photoionization and radiation pressure feedback. This confirms findings from previous works that in low gas surface density environments the effect of radiation feedback is modest.
\item Radiation from stars drastically changes the phase structure of the interstellar medium via photo-heating and photo-electric heating. These heating mechanisms reduce the amount of cold gas in the disc and at the same time increase the mass in the warm phase. The hot phase remains relatively unaffected by radiation as this phase arises mainly due to  SN explosions.
\item The non-equilibrium chemistry module does a good job of reproducing the molecular, atomic, and ionized phases of the gas. Molecular hydrogen is present mainly in the high-density ($1\lesssim n_H \, [\mathrm{cm}^{-3}] \, \lesssim 10^4$) intermediate temperature gas ($100 \lesssim \mathrm{T} \, \mathrm{[K]} \lesssim 10^4$), in agreement with previous works, which allows us to match the Kennicutt-Schmidt relation. 
\item The dust abundances predicted by our simulations match the values inferred for local group galaxies. We show that the dust to gas ratio is a complex function of gas density, temperature, and ionization state, which cannot be captured by simple scaling relations generally used in the literature.  Our dust model is also able to match the scaling relations of dust masses with stellar mass and the atomic gas present in the galaxy.
\item The simulations reproduce observational estimates of the dust temperature distribution of  local group galaxies only if heating from old stellar populations is included, implying that these stars play a  non-negligible role in heating the dust in galaxies. We also show that the median interstellar radiation field heats the dust to temperatures of about $20$~K. The warm dust ($\sim 40$~K) on the other hand lies close to star forming regions and is heated by the action of newly formed O and B stars. 
\end{enumerate}

We have shown that our simulations are capable of reproducing and predicting a wide range of observables. We plan to use this state-of-the-art model to perform next generation cosmological galaxy formation simulations that will be able to predict the resolved ($\sim 10$~pc) properties of galaxies and their outflows with greater fidelity than has been possible previously.  This is an important and necessary step in order to interpret observational data from current and future facilities like {\it ALMA}, {\it JWST}, {\it TMT}, {\it GMT} and {\it E-ELT}, which will provide resolved imaging and spectroscopic data of galaxies all the way up to $z\sim2$.
\section*{Acknowledgements}
RK  acknowledges support from NASA through Einstein Postdoctoral Fellowship
grant number PF7-180163 awarded by the {\it Chandra} X-ray Center, which is
operated by the Smithsonian Astrophysical Observatory for NASA under contract
NAS8-03060. FM is supported by the Program "Rita LEvi Montalcini" of the Italian MIUR. LVS is thankful for the financial support from NASA through HST-AR-14582. Computing resources supporting this work were provided by the NASA High-End Computing (HEC) Program through the NASA Advanced Supercomputing (NAS) Division at Ames Research Center.

\bibliographystyle{mnras}
\bibliography{ISM}

\label{lastpage}
\end{document}